\documentclass[aps,a4]{revtex4}
\usepackage[T1]{fontenc}
\usepackage[latin9]{inputenc}
\usepackage{color}
\usepackage{amsmath}
\usepackage{amssymb}
\usepackage{graphicx}
\usepackage{esint}

\makeatletter

\providecommand{\tabularnewline}{\\}

\@ifundefined{textcolor}{}
{%
 \definecolor{BLACK}{gray}{0}
 \definecolor{WHITE}{gray}{1}
 \definecolor{RED}{rgb}{1,0,0}
 \definecolor{GREEN}{rgb}{0,1,0}
 \definecolor{BLUE}{rgb}{0,0,1}
 \definecolor{CYAN}{cmyk}{1,0,0,0}
 \definecolor{MAGENTA}{cmyk}{0,1,0,0}
 \definecolor{YELLOW}{cmyk}{0,0,1,0}
}

\@ifundefined{definecolor}{\@ifundefined{definecolor}
 {\usepackage{color}}{}
}{}
\@ifundefined{definecolor}{\@ifundefined{definecolor}
 {\@ifundefined{definecolor}
 {\usepackage{color}}{}
}{}
}{}
\@ifundefined{definecolor}{\@ifundefined{definecolor}
 {\@ifundefined{definecolor}
 {\@ifundefined{definecolor}
 {\usepackage{color}}{}
}{}
}{}
}{}
\@ifundefined{definecolor}{\@ifundefined{definecolor}
 {\@ifundefined{definecolor}
 {\@ifundefined{definecolor}
 {\@ifundefined{definecolor}
 {\usepackage{color}}{}
}{}
}{}
}{}
}{}
\usepackage{subfigure}

\makeatother

\begin{document}

\title{Viable cosmological solutions in massive bimetric gravity}

\author{Frank Koennig$^{1}$, Aashay Patil$^{2}$, Luca Amendola$^{1}$}

\affiliation{$^{1}$Institut für Theoretische Physik, Ruprecht-Karls-Universität
Heidelberg, Philosophenweg 16, 69120 Heidelberg, Germany}

\affiliation{$^{2}$Indian Institute of Science Education and Research, Pune-
411008, India}
\begin{abstract}
We find the general conditions for viable cosmological solution at
the background level in bigravity models. Furthermore, we constrain
the parameters by comparing to the Union 2.1 supernovae catalog and
identify, in some cases analytically, the best fit parameter or the
degeneracy curve among pairs of parameters. We point out that a bimetric
model with a single free parameter predicts a simple relation between
the equation of state and the density parameter, fits well the supernovae
data and is a valid and testable alternative to $\Lambda$CD\textcolor{black}{M.
}Additionally, we identify the conditions for a phantom behavior\textcolor{black}{{}
and show that viable bimetric cosmologies cannot cross the phantom
divide.}
\end{abstract}
\maketitle

\section{Introduction}

The discovery of cosmic acceleration has sparked a renewed interest
in theories that go beyond standard gravity. Beside the possibility
of explaining dark energy, the main motivation is to find new observationally
testable features of gravity that allow one to test it beyond the
narrow limits of the solar system.

It is possible to identify three main classes of models of modified
gravity: based on additional scalar fields, vectors or tensors, respectively.
The first one is perhaps the most studied one, owing to the similarity
with inflation and to its simpli\textcolor{black}{city. Even restricting
oneself to single scalar fields with second order equation of motion,
the class of possible Lagrangians, represented by the so-called Horndeski
Lagrangian \cite{Horndeski:1974,Deffayet:2011gz}, is however huge.
In this paper we concern us with the third class, namely models that
modify Einstein's gravity by introducing a massive term in the equations
of motion.}

\textcolor{black}{The history of massive gravity is an old one, dating
back to 1939, when the linear model of Fierz and Pauli was published.
We refer to the review \cite{2012RvMP...84..671H} for a reconstruction
of the steps leading to the modern approach. The key point of these
new forms of massive gravity is the introduction of a second tensor
field, beside the metric. Such a theory of massive gravity was studied
in \cite{2011PhRvL.106w1101D} and was later shown to be free of ghosts
\cite{2012PhRvL.108d1101H}. Furthermore, the interaction of the two
tensor fields creates a mixture of massless and massive gravitons
that apparently avoid the appearance of ghosts \cite{2012JHEP...02..126H}.}

\textcolor{black}{In Ref. \cite{2012JHEP...02..126H,2012JHEP...02..026H}
the authors proposed to render the second tensor field dynamical,
just as the standard metric, although only the latter is coupled to
matter (for a generalization, see \cite{2013JCAP...10..046A}). This
approach, denoted bimetric gravity, keeps the theory ghosts-free and
has the advantage of allowing cosmologically viable solutions. The
cosmology of bimetric gravity has been studied in several papers,
e.g. in Refs. \cite{2013JHEP...03..099A,Comelli:2012db,2013arXiv1304.3920D,2012JHEP...03..067C,2012JHEP...01..035V,1475-7516-2012-03-042}.
The main conclusion is that bimetric gravity allows for a cosmological
evolution that can approximate the $\Lambda$CDM universe and can
therefore be a candidate for dark energy. For a criticism of these
theories see e.g. Ref. \cite{2013PhRvD..88h1501D}, whose conclusions
are however apparently contradicted by the results in \cite{2013arXiv1303.6940H}.}

\textcolor{black}{Bimetric gravity has been compared to background
data, in particular supernovae Ia, in \cite{2013JHEP...03..099A,1475-7516-2012-03-042},
where confidence regions have been obtained for various cases. We
will recover indeed several results already presented in \cite{2013JHEP...03..099A}.
We feel however that several interesting questions concerning the
possibility of obtaining a viable cosmological evolution in bimetric
models have not been fully addressed yet. Some of the questions that
this paper addresses are: 1) For which values of the parameters and
of the initial conditions does bimetric gravity allow for viable cosmologies?
2) For which values of the parameter there appear an ef}fective phantom
(i.e. an equation of state less than -1) behavior? 3) Can one find
simple expressions for the parameters for which the supernovae data
can be fitted?

We will find that in several cases these questions can be answered
in a simple analytical way, providing a number of alternatives to
$\Lambda$CDM\textcolor{black}{. Interestingly, these alternative
models do not reduce to $\Lambda$CDM for some values of the parameters
(unless of course a cosmological constant is added as an additional
parameter) and can therefore be ruled out by precise cosmological
observations (if they are not yet ruled out!). }In particular, we
point out that a minimal bimetric model\emph{ }with a single free
parameter predicts a simple relation between the equation of state
and the density parameter, fits well the supernovae data and is a
valid and testable alternative to $\Lambda$CDM.

\textcolor{black}{The results of this paper provide a preliminary
choice of well-behaved cos}mological evolutions that can be further
analyzed at the perturbation level. This task will be carried out
in a companion paper.

\section{Background equations}

We start with the action of the form \cite{2012JHEP...02..126H} 
\begin{eqnarray}
S & = & -\dfrac{M_{g}^{2}}{2}\int d^{4}x\,\sqrt{-\det g}\, R(g)-\dfrac{M_{f}^{2}}{2}\int d^{4}x\,\sqrt{-\det f}\, R(f)\\
 & + & m^{2}M_{g}^{2}\int d^{4}x\,\sqrt{-\det g}\,\sum_{n=0}^{4}\beta_{n}e_{n}\left(\sqrt{g^{\alpha\beta}f_{\beta\gamma}}\right)+\int d^{4}x\,\sqrt{-\det g}\, L_{m}(g,\Phi)
\end{eqnarray}
where $e_{n}$ are suitable polynomials and $\beta_{n}$ arbitrary
constants. Here $g_{\mu\nu}$ is the standard metric coupled to matter
fields in the $L_{m}$ Lagrangian, while $f_{\mu\nu}$ is a new dynamical
tensor field. In the following we express masses in units of $M_{g}^{2}$
and the mass parameters $m^{2}$ will be absorbed into the parameters
$\beta_{n}$. The action then becomes 
\begin{eqnarray}
S & = & -\dfrac{1}{2}\int d^{4}x\,\sqrt{-\det g}\, R(g)-\dfrac{M_{f}^{2}}{2}\int d^{4}x\,\sqrt{-\det f}\, R(f)\\
 & + & \int d^{4}x\,\sqrt{-\det g}\,\sum_{n=0}^{4}\beta_{n}e_{n}\left(\sqrt{g^{\alpha\beta}f_{\beta\gamma}}\right)+\int d^{4}x\,\sqrt{-\det g}\, L_{m}(g,\Phi)\;.
\end{eqnarray}
Varying the action with respect to $g{}_{\mu\nu}$, one obtains the
following equations of motion, 
\begin{equation}
R_{\mu\nu}-\dfrac{1}{2}g_{\mu\nu}R+\dfrac{1}{2}\sum_{n=0}^{3}(-1)^{n}\beta_{n}\left[g_{\mu\lambda}Y_{(n)\nu}^{\lambda}\left(\sqrt{g^{\alpha\beta}f_{\beta\gamma}}\right)+g_{\nu\lambda}Y_{(n)\mu}^{\lambda}\left(\sqrt{g^{\alpha\beta}f_{\beta\gamma}}\right)\right]=T_{\mu\nu}\label{eq:eeg}
\end{equation}
where the expressions $Y_{(n)\nu}^{\lambda}\left(\sqrt{g^{\alpha\beta}f_{\beta\gamma}}\right)$
are defined as, putting $X=\left(\sqrt{g^{-1}f}\right)$, 
\begin{align}
Y_{(0)}(X) & =I,\\
Y_{(1)}(X) & =X-I[X],\\
Y_{(2)}(X) & =X^{2}-X[X]+\dfrac{1}{2}I\left([X]^{2}-[X^{2}]\right)\\
Y_{(3)}(X) & =X^{3}-X^{2}[X]+\dfrac{1}{2}X\left([X]^{2}-[X^{2}]\right)-\dfrac{1}{6}I\left([X]^{3}-3[X][X^{2}]+2[X^{3}]\right)
\end{align}
where $I$ is the identity matrix and $[...]$ is the trace operator.

Varying the action with respect to $f{}_{\mu\nu}$ we get 
\begin{equation}
\bar{R}_{\mu\nu}-\dfrac{1}{2}f_{\mu\nu}\bar{R}+\dfrac{1}{2M_{f}^{2}}\sum_{n=0}^{3}(-1)^{n}\beta_{4-n}\left[f_{\mu\lambda}Y_{(n)\nu}^{\lambda}\left(\sqrt{f^{\alpha\beta}g_{\beta\gamma}}\right)+f_{\nu\lambda}Y_{(n)\mu}^{\lambda}\left(\sqrt{f^{\alpha\beta}g_{\beta\gamma}}\right)\right]=0
\end{equation}
where the overbar indicates $f{}_{\mu\nu}$ curvatures. Under the
rescaling $f\rightarrow M_{f}^{-2}f$, the Ricci scalar transforms
as $\bar{R}(f)\rightarrow M_{f}^{2}\bar{R}(f)$ which results in 
\begin{equation}
\sqrt{-\det f}\bar{R}(f)\rightarrow M_{f}^{-2}\sqrt{-\det f}\bar{R}(f)\;.
\end{equation}
Next to the Einstein-Hilbert term for $f_{\mu\nu},$ there is another
term in the action that depends on $f_{\mu\nu}$ which transforms
as 
\begin{equation}
\sum_{n=0}^{4}\beta_{n}e_{n}\left(\sqrt{g^{-1}f}\right)\rightarrow\sum_{n=0}^{4}\beta_{n}e_{n}\left(M_{f}^{-1}\sqrt{g^{-1}f}\right)\;.
\end{equation}
Since the elementary symmetric polynomials $e_{n}(X)$ are of order
$X^{n}$, the rescaling of $f_{\mu\nu}$ by a constant factor $M_{f}^{-2}$
translates into a redefinition of the couplings $\beta_{n}\rightarrow M_{f}^{n}\beta_{n}$
which allows us to assume $M_{f}=1$ in the following.

We assume now a cosmological spatially flat FRW metric 
\begin{equation}
ds^{2}=a^{2}(t)\left(-dt^{2}+dx_{i}dx^{i}\right)
\end{equation}
where $t$ represents the conformal time and a dot will represent
the derivative with respect to it. The second metric is chosen as
\begin{equation}
f_{\mu\nu}=\begin{bmatrix}-\dot{b}(t)^{2}/\mathcal{H}^{2}(t) & 0 & 0 & 0\\
 & b(t)^{2}\\
0 & 0 & b(t)^{2} & 0\\
0 & 0 & 0 & b(t)^{2}
\end{bmatrix}
\end{equation}
where $\mathcal{H}\equiv\dot{a}/a$ is the conformal Hubble function.
This form of the metric $f_{\mu\nu}$ ensures that the equations satisfy
the Bianchi identities (see e.g. \cite{2012JHEP...02..026H}).

Inserting $g_{\mu\nu}$ in Eq. (\ref{eq:eeg}) we get 
\begin{eqnarray}
3\mathcal{H}{}^{2} & = & a^{2}\left(\rho_{m}+\rho_{mg}\right)\label{eq:fried-1}
\end{eqnarray}
where the massive gravity energy density is 
\begin{equation}
\rho_{mg}=B_{0}\equiv\beta_{0}+3\beta_{1}r+3\beta_{2}r^{2}+\beta_{3}r^{3}\label{eq:rho_mg}
\end{equation}
with 
\begin{equation}
r=\frac{b}{a}
\end{equation}
The matter energy density follows the usual conservation law 
\begin{equation}
\dot{\rho}_{m}+3\mathcal{H}\rho_{m}=0\;.\label{eq:mattcons}
\end{equation}
\textcolor{black}{Notice that although we do not consider explicitly
a radiation epoch (since we confine ourselves to observations at low
redshifts), a radiation component could be easily added to the pressureless
matter and would not change qualitatively any of the conclusions below.
W}e can also define 
\begin{equation}
\Omega_{mg}=\frac{\rho_{mg}}{\rho_{m}+\rho_{mg}}=1-\Omega_{m}\label{eq:Om_mg}
\end{equation}
where $\Omega_{m}=\rho_{m}/\left(\rho_{m}+\rho_{mg}\right)$.

Similarly, the background equation for the $f$ metric is 
\begin{equation}
\mathcal{H}^{2}=\frac{a^{2}}{3r}B_{1}\label{eq:frif}
\end{equation}
if $B_{1}\not=0$ (a\textcolor{black}{nd $\dot{b}=0$ if $B_{1}=0$)
wh}ere 
\begin{equation}
B_{1}=\beta_{1}+3\beta_{2}r+3\beta_{3}r^{2}+\beta_{4}r^{3}\;.
\end{equation}
Combining (\ref{eq:fried-1}) and (\ref{eq:frif}), differentiating
and inserting (\ref{eq:mattcons}) we obtain the constraint 
\begin{equation}
\dot{b}=-\frac{\left(4\beta_{0}+9\beta_{1}r+6\beta_{2}r+\beta_{3}r^{2}\right)\dot{a}}{3B_{2}}\label{eq:bdot}
\end{equation}
where 
\begin{equation}
B_{2}=\beta_{1}+2\beta_{2}r+\beta_{3}r^{2}\;.
\end{equation}
The background equations can be conveniently written as a first ord\textcolor{black}{er
system for $r$ and $\mathcal{H}$, where the prime denotes the derivative
with respect to $N=\log a$:} 
\begin{align}
2\mathcal{H}'\mathcal{H}+\mathcal{H}^{2} & =a^{2}\left(B_{0}+B_{2}r'\right)\;,\label{eq:Hprime}\\
r' & =\frac{3rB_{1}\Omega_{m}}{\beta_{1}-3\beta_{3}r^{2}-2\beta_{4}r^{3}+3B_{2}r^{2}}\;,\label{eq:rprime}\\
\Omega_{m} & =1-\frac{B_{0}}{B_{1}}r\label{eq:omegam}
\end{align}
(the $r'$ equation has been first obtained in Ref. \cite{2013JHEP...03..099A}).
We can define the effective equation of state 
\begin{align}
w_{eff} & \equiv\Omega_{mg}w_{mg}=-\frac{1}{3}\left(1+2\frac{\mathcal{H}'}{\mathcal{H}}\right)=-\frac{r\left(B_{0}+B_{2}r'\right)}{B_{1}}\;,\label{eq:w_eff_def}\\
 & =-1+\Omega_{m}-\frac{B_{2}rr'}{B_{1}}\;.\label{eq:w_eff}
\end{align}
Eq. (\ref{eq:rprime}) is particularly useful for our discussion below.
Notice that it can be written also as 
\begin{equation}
r'=-\frac{3\rho_{m}}{\rho_{m,r}}\label{eq:rprimerho}
\end{equation}
where $\rho_{m,r}$ denotes differentiation with respect to $r$ of
the function 
\begin{equation}
\rho_{m}(r)=\frac{B_{1}}{r}-B_{0}\label{eq:rho_m}
\end{equation}
obtained by combining Eqs. (\ref{eq:fried-1}) and (\ref{eq:frif}).

It is convenient from now on to express the $\beta$ parameters in
units of $H_{0}^{2}$ and $H$ in units of $H_{0}$.

\section{Conditions for cosmological viability}

Several possible branches of the solution of Eq. (\ref{eq:rprime})
are possible, depending on the initial condition for\textcolor{black}{{}
$r$. We distinguish in the following between }\textcolor{black}{\emph{finite}}\textcolor{black}{{}
branches, that are confined within two successive roots or poles of
$r'$, and }\textcolor{black}{\emph{infinite}}\textcolor{black}{{}
branches, which can extend to infinite values of $r$. We define now
a viable cosmological solution one in which the following conditions
are satisfied: $a)$ $\rho_{m}>0$ }and $\rho_{mg}$ not identically
zero,\textcolor{black}{{} }\textcolor{black}{\emph{b}}\textcolor{black}{)}
a monotonic expansion, i.e. $\rho_{m}+\rho_{mg}>0$,\textcolor{black}{{}
}\textcolor{black}{\emph{c}}\textcolor{black}{) the evolution in the
}asymptotic past is dominated by $\rho_{m}$, i.e. \textcolor{black}{$\rho_{m}(N\to-\infty)\rightarrow\infty$,}
$\Omega_{m}(N\to-\infty)=1$ (and therefore $w_{eff}(N\rightarrow-\infty)=0$),
\emph{d}) no singularities in $r'$ at finite times and \emph{e})
$r\ge0$ at all ti\textcolor{black}{mes. Violations of these conditions
do not necessarily imply contradiction with observable data if they
occur outside the observable range and could in principle be lifted
or relaxed. However, when they are satisfied the cosmological evolution
is much safer, simpler and requires no special tuning. Most of what
follows is devoted to determining the conditions under which cosmological
viable solutions take place.}

Combining these conditions and analyzing Eq. (\ref{eq:rprime}) yields
the following properties of viable models: 
\begin{enumerate}
\item All viable models except $\beta_{i}=0\,\forall\, i>0$, i.e. the $\Lambda$CDM
case, must fulfill $r'\rightarrow-\infty$ as $r\rightarrow\infty$.
To see this, we use Eq. (\ref{eq:rprime}) to find that models in
which we can not observe this limit need to satisfy $\beta_{1}=\beta_{3}=0$
and $\beta_{2}=\frac{1}{3}\beta_{4}$. With this choice, the combination
of Eq. (\ref{eq:Hprime}) and the background equation (\ref{eq:frif})
together with its derivative yields 
\begin{equation}
\sqrt{\beta_{2}\left(1+r^{2}\right)}\left(\beta_{0}-3\beta_{2}\right)=0
\end{equation}
which provides\textcolor{black}{{} the constraint $\beta_{0}=3\beta_{2}$.
But this corresponds to a vanishing matter density $\rho_{m}$ which
is not viable. Note that the choice of parameters $\beta_{1}=\beta_{3}=0$
and $\beta_{0}=3\beta_{2}=\beta_{4}$ matches with those of the partially
massless bimetric theory which was studied in \cite{2013PhLB..726..834H}.
However, in those theories the authors assumed the reference metric
to be proportional to $g_{\mu\nu}$ which is explicitly avoided in
this work due our choice of the Bianchi constraint.}
\item \label{ point2}If a viable range in $r$ is infinite th\textcolor{black}{en,
as just shown, $r$ decreases with time sin}ce the limit $r'\rightarrow-\infty$
as $r\rightarrow\infty$ must hold. Then $r\to\infty$ corresponds
to the infinite past and therefore, if this branch is viable, then
it needs to satisfy $\lim_{r\rightarrow\infty}\Omega_{m}=1$. With
Eq. (\ref{eq:omegam}) one finds that a viable solution with an infinite
range in $r$ requires $\beta_{2}=\beta_{3}=0\neq\beta_{4}$. Moreover,
$\beta_{4}$ is enforced to be positive in order to produce a positive
expansion rate at early times. 
\item A non-vanishing massive gravity part, i.e. $B_{0}\neq0$, always implies
that if there is a root $r=0$, then for this root, and only for this
one, $\Omega_{m}=1$. For all other roots we need $\Omega_{m}=0$
in order to fulfill Eq. (\ref{eq:rprime}). 
\item Let $r\in(r_{1},r_{2})$ be a bra\textcolor{black}{nch with $r'|_{r_{1}}=r'|_{r_{2}}=0$
for $r_{1},r_{2}$ strictly positive. As seen before, a root at $r>0$
corresponds to $\rho_{m}=0$. For a non-constant evolution of the
matter density, the mean value theorem always provides a $\bar{r}\in(r_{1},r_{2})$
with $\rho_{m,r}=0$ causing a singularity in $r'$.}\textcolor{red}{{}
} Since Eq. (\ref{eq:rho_m}) shows that the matter density can not
become divergent at a finite and non-zero $r$, a{} viable model
always evolves from either $r=0$ or $r=\infty$ to a root of $\rho$\label{enu:4}.
\item We will find that $r=0$ always corresponds to the asymptotic past.
If it would instead describe a final state, then a vanishing $\rho_{m}$
as $N\rightarrow\infty$ (which has to hold since the matter density
follows the usual conservation rule) needs $\beta_{1}=0$ and $\beta_{0}=3\beta_{2}$.
Additionally, this requires $\beta_{\text{3}}>0$, otherwise we have
either a negative $\beta_{3}$ which means that the density is not
positive or $\beta_{3}=0$ in which the branch would be infinite with
$\rho_{mg}=0$, i.e. $\Omega_{m}=1$, at all times. However, we then
obtain a finite branch between two roots of $\rho(r)$ at $r=0$ and
$r_{c}>0$ but we already concluded in point \ref{enu:4} that $r=0$
must then correspond to the asymptotic past. 
\item The previous conclusions imply for all viable cases an evolution from
$\Omega_{m}=1$ to the final state $\Omega_{m}=0$, just like $\Lambda$CDM. 
\item We can use Eq. (\ref{eq:rprime}) to find that there is always a root
at $r=0$ for non-vanishing $\beta_{1}$. All models without a root
at $r=0$ need to satisfy 
\begin{equation}
\lim_{r\rightarrow0}\Omega_{m}\big|_{\beta_{1}=0}=1-\lim_{r\rightarrow0}\frac{\beta_{0}+3\beta_{2}r^{2}+\beta_{3}r^{3}}{3\left(\beta_{2}+\beta_{3}r\right)+\beta_{4}r^{2}}=1
\end{equation}
In this case, viability enforces $\beta_{0}=0.$ Models with a pole
at $r=0$ need to satisfy $\beta_{1}=\beta_{3}=0$ with $\beta_{2}\neq\frac{1}{3}\beta_{0}$
and must fulfill 
\begin{equation}
\lim_{r\rightarrow0}w_{eff}\big|_{\beta_{0}=\beta_{1}=\beta_{3}=0}=-\frac{3\beta_{2}}{3\beta_{2}-\beta_{4}}=0\;.
\end{equation}
This contradicts the condition $\beta_{2}\neq0$. If $r=0$ is neither
a root nor a pole, then from Eq. (\ref{eq:rprime}) we see that this
corresponds to $\beta_{3}\neq0$ and $\beta_{0}\neq3\beta_{2}$ (note
that this implies $\beta_{2}\neq0$) instead. However, the resulting
matter density 
\begin{equation}
\rho_{m}\Big|_{\beta_{0}=\beta_{1}=0}=3\beta_{2}+3\beta_{3}r+\left(\beta_{4}-3\beta_{2}\right)r^{2}-\beta_{3}r^{3}
\end{equation}
violates the requirement of a divergent density for $r\rightarrow0$.
Therefore, every viable branch that evolves from $r=0$ must satisfy
$r'|_{r=0}=0$. 
\item If $r$ evolves from $r=0$, then a positive $H^{2}$ at early times
implies $\beta_{k}>0$ where $\beta_{k}$ denotes the non-vanishing
$\beta$-parameter with the smallest index $k\neq0$. 
\item A model which produces two viable branches has to satisfy $\beta_{1}\geq0$
and $\beta_{4}>0$, in order to produce positive Hubble functions
in both branches. 
\item From Eq. (\ref{eq:w_eff}) we find that the equation of state always
evolves from $w_{eff}=0$, as required from the conditions of viability,
to $w_{eff}=-1$ on a viable solution. Notice that $w_{eff}=-1$ even
for a vanishing explicit cosmological constant $\beta_{0}=0$.\\
 \\
 Depending on the n\textcolor{black}{umber of non-negative roots,
we therefore find that several cases can not be viable: } \\

\item \textcolor{black}{The number of non-negative roots can be zero only
if $\beta_{1}=\beta_{2}=\beta_{3}=0$, which leads to 
\begin{equation}
r'=\frac{3\left(\beta_{0}-\beta_{4}r^{2}\right)}{2\beta_{4}r}\;.
\end{equation}
As already remarked, a viable model must therefore evolve from $r=\infty$
to $r=0$ since $r'<0$ for $r\rightarrow\infty$ and this requires
a positive and non-zero $\beta_{4}$. However, this produces a }singular
$r'$ at $r=0$\textcolor{black}{{} (unless $\beta_{0}=0$ but we
are now on}ly interested in models with no positive roots) which was
already shown to be non-viable. 
\item A model that has at least one \textcolor{black}{positive} root and
does not have a root at $r=0$ may only be able to produce a viable
infinite branch. A finite but non-zero $r'$ at $r=0$ can not be
achieved with a vanishing $\beta_{3}$ but this is enforced by the
criteria of viable infinite branches (see point \ref{ point2}). Thus,
all models with only non-zero roots must fulfill $\beta_{1}=\beta_{2}=\beta_{3}=0<\beta_{4},\beta_{0}$.
With e.g. Descartes' rule of sign we see that we can not expect more
than one positive root. Whenever there is a model with at least two
\textcolor{black}{positive} roots producing a viable branch, there
must be one root at $r=0$. 
\item If there is only \textcolor{black}{one} root $r=0$, then this root
is reached in the asymptotic future, i.e. for $N=\infty$, since the
range must be infinite. This contradicts the previous conclusion that
$r=0$ has to correspond to the asymptotic past. Therefore, no viable
cosmologies exist if there is only one root at $r=0$.\label{enu:11} 
\item If there are $n\geq2$ positive roots at $r_{c_{1}},...,r_{c_{n}}$,
where $r_{c_{i}}<r_{c_{j}}$for $i<j$, then only the two branches
$r\in(0,r_{c_{1}})$ and $r\in(r_{c_{n}},\infty)$ may be viable. 
\item Models with two viable branches require $\beta_{2}=\beta_{3}=0$ and
$\beta_{1},\beta_{4}>0$. Descartes' rule of sign then shows that
those models must have exactly two positive roots. 
\item With, again, Descartes' rule of sign we find that there is no model
with $\beta_{2}=\beta_{3}=0$ that produce three positive roots. For
this reason, we can not expect any viable infinite branch in models
with three positive roots. 
\end{enumerate}
Finally, we can employ these results to show that se\textcolor{black}{veral
simple models do not produce viable solutions: } 
\begin{itemize}
\item \textcolor{black}{Consider models in which only one $\beta$-parameter
does not vanish. Let's call them $\beta_{i}$ models. Then only $\beta_{0}$
or $\beta_{1}$ models may produce viable solutions. This first one
is not surprising since it is equivalent to a $\Lambda$CDM universe.
For all the other $\beta_{i}$ models, we find 
\begin{equation}
r'\Big|_{\beta_{i}=0,i\neq2}=-\frac{3\left(r^{2}-1\right)}{2r}\;,\qquad r'\Big|_{\beta_{i}=0,i\neq3}=-\frac{r\left(r^{2}-3\right)}{r^{2}-1}\;,\qquad r'\Big|_{\beta_{i}=0,i\neq4}=-\frac{3}{2}r\;.
\end{equation}
The infinite branch in $\beta_{2}$ or $\beta_{3}$ models can not
be viable. In addition, their finite branches suffer from a pole in
$r'$. Therefore, we can not expect any viable solutions. These arguments
do not hold for the $\beta_{4}$ model. However, we already concluded}
(see point \textcolor{black}{\ref{enu:11}) that a model with only
one root at $r=0$ is not viable. } 
\item \textcolor{black}{In a more general case, in which two free $\beta$-parameters
are allowed to vary (let's denote them $\beta_{i}\beta_{j}$ models),
we will find that only the combination involving $\beta_{0}$ or $\beta_{1}$
are generally able to produce viable solutions.} To see that the models
$\beta_{2}\beta_{3}$, $\beta_{2}\beta_{4}$ and $\beta_{3}\beta_{4}$
can not be viable, we first assume that both couplings in all three
combinations do not vanish, otherwise we would obtain non-viable minimal
models. \textcolor{black}{This also rejects the possibility of viable
models with an infinite branch in these cases. In the $\beta_{2}\beta_{3}$
model, the matter density evolves like 
\begin{equation}
\rho_{m}=3\left(\beta_{2}+\beta_{3}r-\beta_{2}r^{2}\right)-\beta_{3}r^{3}\;,
\end{equation}
} and is therefore finite at $r=0$, \textcolor{black}{which contradicts
condition $c)$. In fact this solution can be continued to negative
$r$, which implies that $|b|$ reaches zero and increases again.
This is therefore a bouncing cosmology which is interesting on its
own but violates our viability condition and we leave its study to
future work}\textcolor{blue}{.}\textcolor{black}{{} For the $\beta_{2}\beta_{4}$
model we have already shown that only a finite branch ($0,r_{c}$)
could be viable. Simplifying Eq. (\ref{eq:rprime}) yields 
\begin{equation}
r'\Big|_{\beta_{i}=0,i\neq2,4}=-\frac{3}{2}r+\frac{3\beta_{2}}{2r\left(3\beta_{2}-\beta_{4}\right)}\;.
\end{equation}
This exhibits a pole at $r=0$ which indicates non-viability. To analyze
the $\beta_{3}\beta_{4}$ models, we again use Eq. (\ref{eq:rprime})
which directly shows that we need to have $\beta_{3}\neq0$ in order
to get a positive root. In this case, the only positive root is given
by 
\begin{equation}
r_{c}=\frac{\beta_{4}+\sqrt{12+\beta_{3}^{2}+\beta_{4}^{2}}}{2\beta_{3}}\;.
\end{equation}
In addition, we will find that $r'$ is singular at 
\begin{equation}
r_{s}=\frac{\beta_{4}+\sqrt{9\beta_{3}^{2}+\beta_{4}^{2}}}{3\beta_{3}}\;.
\end{equation}
Since $\beta_{3}\neq0$, only the branch (0,$r_{c}$) could be viable
and therefore either $r_{s}<0$ or $r_{s}>r_{c}$ must hold. Notice
that $r_{s}=0$ is not viable.}\textbf{\textcolor{black}{{} }}\textcolor{black}{Both
relations require $\beta_{3}<0$. However, a positive Hubble function
enforces $\beta_{3}>0$ which shows that the branch ($0,r_{c}$) always
contains a singularity in $r'$. We therefore conclude that models
with $\beta_{0}=\beta_{1}=\beta_{2}=0$ are not able to produce viable
solutions. } 
\item \textcolor{black}{The subset of cosmological solutions with an infinite
range in $r$ and without an explicit cosmological constant is described
by the relation $\beta_{0}=\beta_{2}=\beta_{3}=0<\beta_{4}$ together
with $\beta_{1}\neq0$. For these models, we obtain 
\begin{equation}
\Omega_{m,r}=\frac{3\beta_{1}r\left(-2\beta_{1}+\beta_{4}r^{3}\right)}{\left(\beta_{1}+\beta_{4}r^{3}\right)^{2}}
\end{equation}
from which we see that $\Omega_{m}$ increases with time when the
following condition holds: 
\begin{equation}
\Omega_{m,r}<0\;\Longleftrightarrow\text{\;}\left(\beta_{1}<0\;\wedge\;\beta_{1}+\beta_{4}r^{3}\neq0\right)\;\vee\;\left(\beta_{1}>0\;\wedge\; r<\left(\frac{2\beta_{1}}{\beta_{4}}\right){}^{\frac{1}{3}}\right)\;.
\end{equation}
Viable models are therefore only possible if $\beta_{1}>0$. In addition,
the solution $r_{c}$ of the equation 
\begin{equation}
\Omega_{m}=1-\frac{3\beta_{1}r_{c}^{2}}{\beta_{1}+\beta_{4}r_{c}^{3}}=0
\end{equation}
is negative (or zero but this, as already discussed, does not correspond
to a viable solution) if $\beta_{4}>2\beta_{1}$. This shows that
only models with $\beta_{0}=0$ satisfying $\beta_{0}=\beta_{2}=\beta_{3}=0<\frac{1}{2}\beta_{4}\leq\beta_{1}$
are able to produce viable branches $(r_{c},\infty)$.} 
\item A simple model with all identical couplings, i.e. $\beta_{0}=\beta_{i}=\hat{\beta}$,
needs $\hat{\beta}>0$ in order to produce a positive expansion rate.
The matter density\textcolor{black}{{} 
\begin{equation}
\rho_{m}=-\hat{\beta}\frac{(r-1)(r+1)^{3}}{r}
\end{equation}
then shows that only the finite branch (0,$r_{c}$) with $r_{c}=1$
could be viable. Additionally, the Hubble function at present time
\begin{equation}
\frac{1}{3}\hat{\beta}\left(1+r_{0}\right){}^{3}r_{0}^{-1}=1
\end{equation}
is only solved by a purely real and positive present value $r_{0}$
if $\hat{\beta}\leq\frac{4}{9}$. } 
\end{itemize}
\textcolor{black}{In practice, to see if a viable solution exists,
one first has to find all positive solutions $r_{0}$ that fulfill
both Friedmann equations (\ref{eq:fried-1}) and (\ref{eq:frif})
at present time. }One then needs to check \textcolor{black}{whether
the branches $r\in(0,r_{c_{1}})$ and $r\in(r_{c_{n}},\infty)$, where}
\textcolor{black}{$r_{c_{1}}$ and $r_{c_{n}}$ denote the smallest
and largest strictly positive root of $\rho_{m}(r)$, respectively,
contain $r_{0}$ and, finally, ensure that those branches do not contradict
the criteria of viability.}\textcolor{red}{{} }In general, one can
show that a finite branch between two roots $(0,r_{c})$ with $0<r_{0}<r_{c}$
in which $r'$ is positive and does not have any pole is always viable
if the matter density is positive in this range. This provides a very
simple recipe to find viable cosmologies without solving the evolution
equations.

\textcolor{black}{It is also interesting to provide the general conditions
for a phantom ($w_{mg}<-1$) solution to appear. From $w_{eff}$ we
see that 
\begin{equation}
w_{mg}=-1-\frac{B_{2}rr'}{\Omega_{mg}B_{1}}\;.
\end{equation}
Combining with Eq. (\ref{eq:omegam}) we obtain 
\begin{equation}
w_{mg}=-1-\frac{B_{2}}{B_{0}}r'\;.\label{eq:wmg}
\end{equation}
Near the de Sitter final state we can assume $\Omega_{m}\to0$ and
therefore $B_{0}=B_{1}/r$ from Eq. (\ref{eq:omegam}). This implies
\begin{equation}
w_{mg}\approx-1-\frac{B_{2}}{B_{1}}rr'\;.\label{eq:wmg_deSitter}
\end{equation}
In a viable branch with a finite range in $r$, both $r$ and $r'$
are positive. If the range is infinite, then $r'$ is negative. In
addition, $B_{1}$ is always positive due to Eq. (\ref{eq:frif}).
We conclude that a necessary and sufficient condition for a phantom
}equation of state\textcolor{black}{{} is $B_{2}>0$ for a finite
branch $(0,r_{c_{1}})$. If the branch is infinite, then a phantom
requires $B_{2}<0$ which results in $\beta_{1}<0$ since viable models
in infinite branches need to fulfill $\beta_{2}=\beta_{3}=0$. From
Eq. (\ref{eq:bdot}) we notice that $B_{2}$ cannot be zero in a viable
region of $r$ and therefore $w_{mg}$ cannot cross the $-1$ line.
This shows that every viable bigravity cosmology is either phantom
or non-phantom throughout its evolution. Conversely, finding a phantom
crossing would rule out the entire class of viable bimetric cosmologies.}

\begin{figure}
\includegraphics[width=0.45\textwidth]{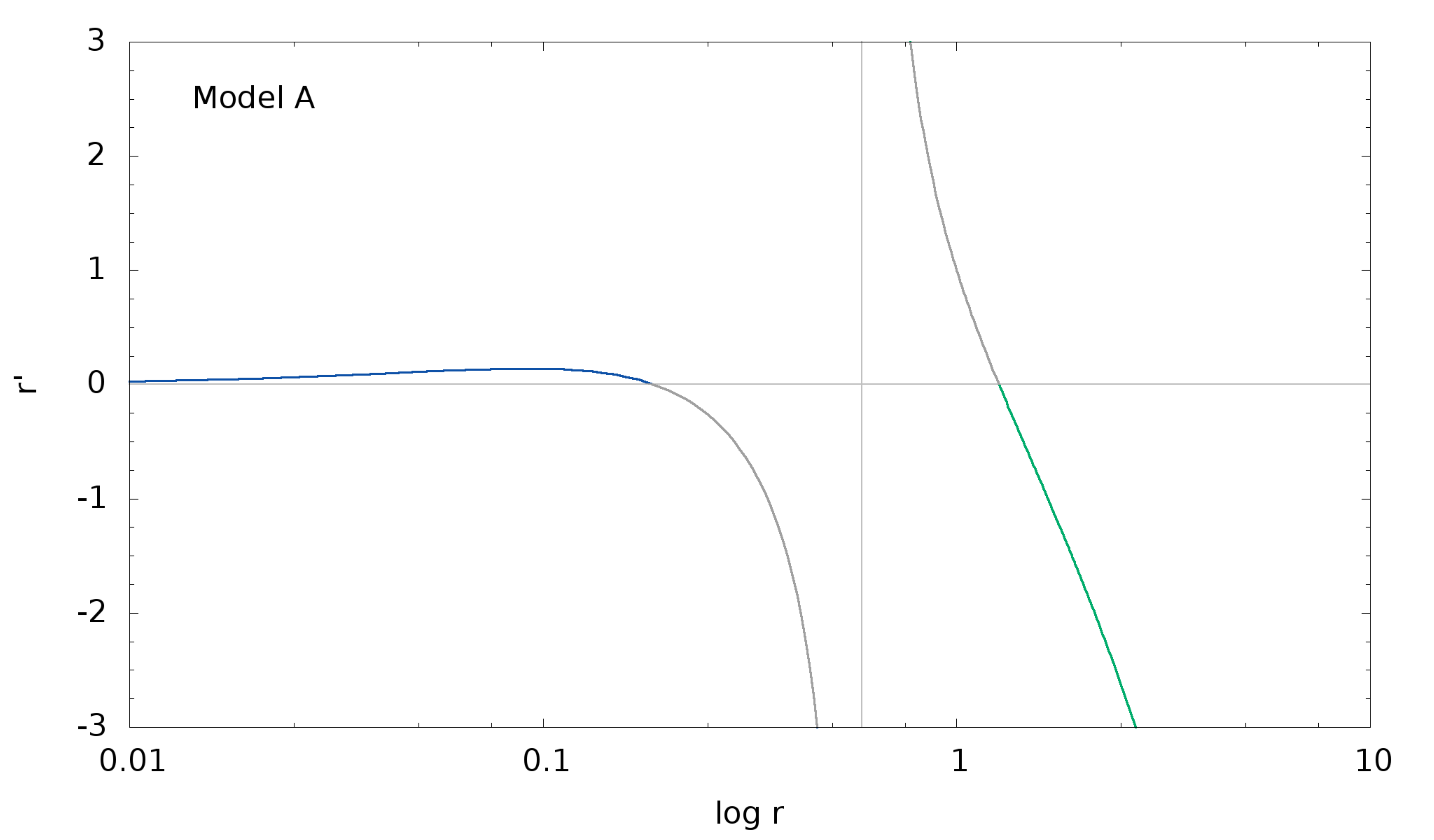}\includegraphics[width=0.45\textwidth]{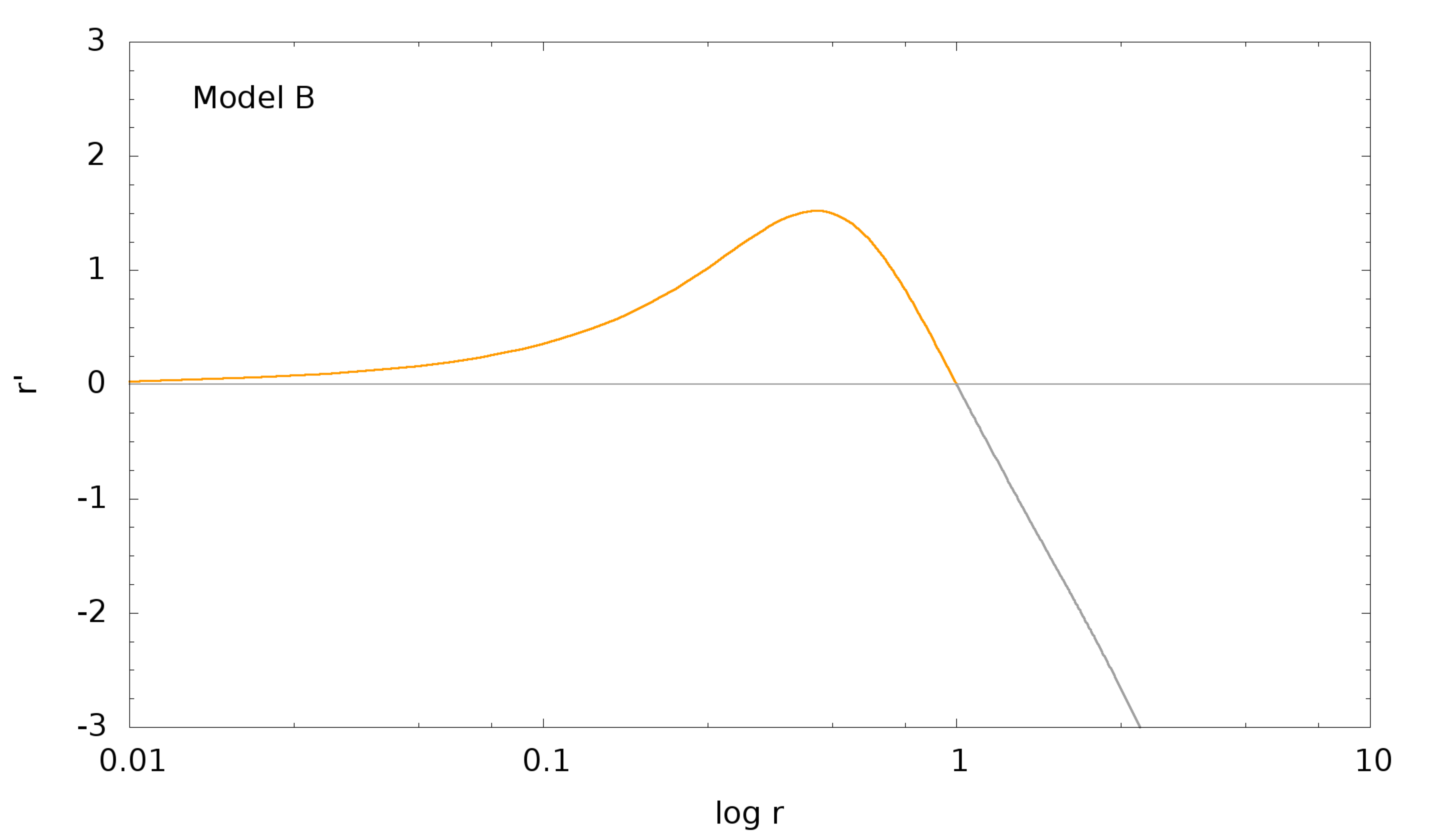}

\caption{The evolution of $r'(r)$ corresponding to the models A and B with
$\beta_{i}=(1,\frac{1}{5},0,0,1)$ (left) and $\beta_{i}=\frac{4}{9}$
(right), respectively, visualizing all possible branches. The first
model contains two finite ($\sim(0,0.2)$ and $\sim(0.2,1.30)$) and
one infinite branch ($\sim(1.3,\infty)$). However, only the first
and third branch may be viable which, indeed, turns out to be the
case. On the contrary, the one-parameter model B only produces one
viable branch $(0,1)$ with $r_{0}=\frac{1}{2}$, though $r'$ seems
to evolve viable even in the infinite branch. \label{fig:example_models_rp}}
\end{figure}

\begin{figure}
\includegraphics[width=0.45\textwidth]{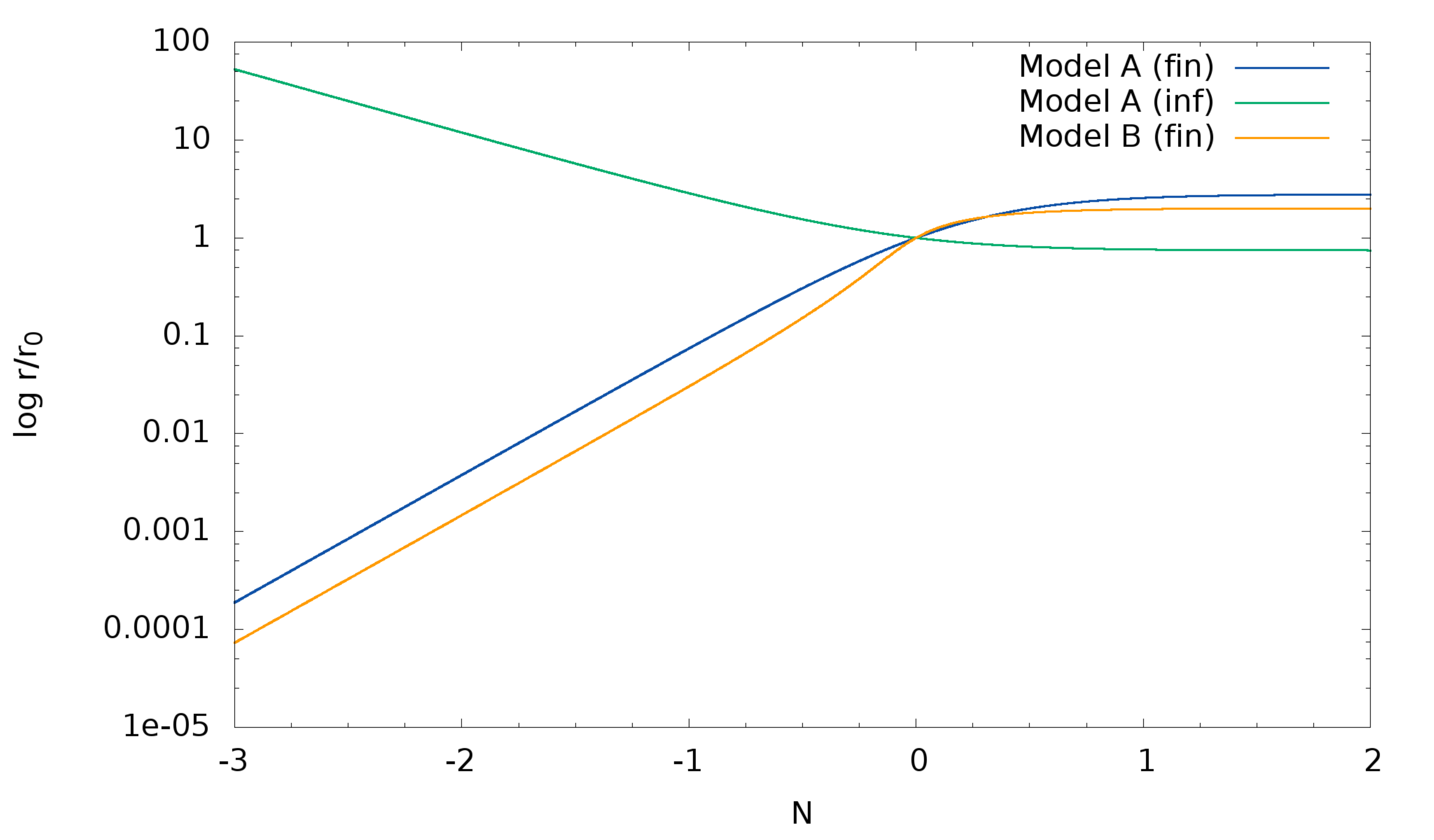}\includegraphics[width=0.45\textwidth]{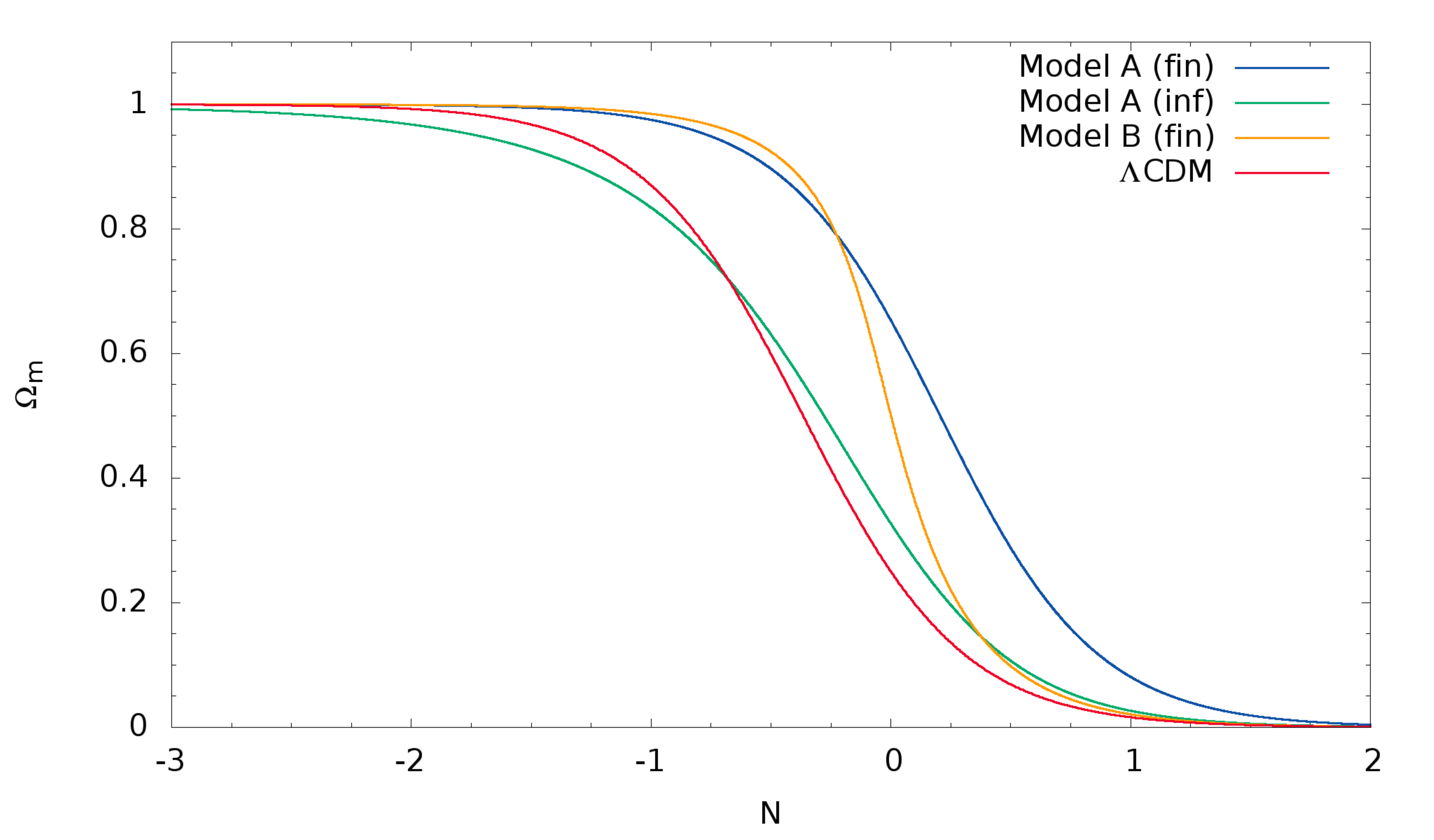}

\includegraphics[width=0.45\textwidth]{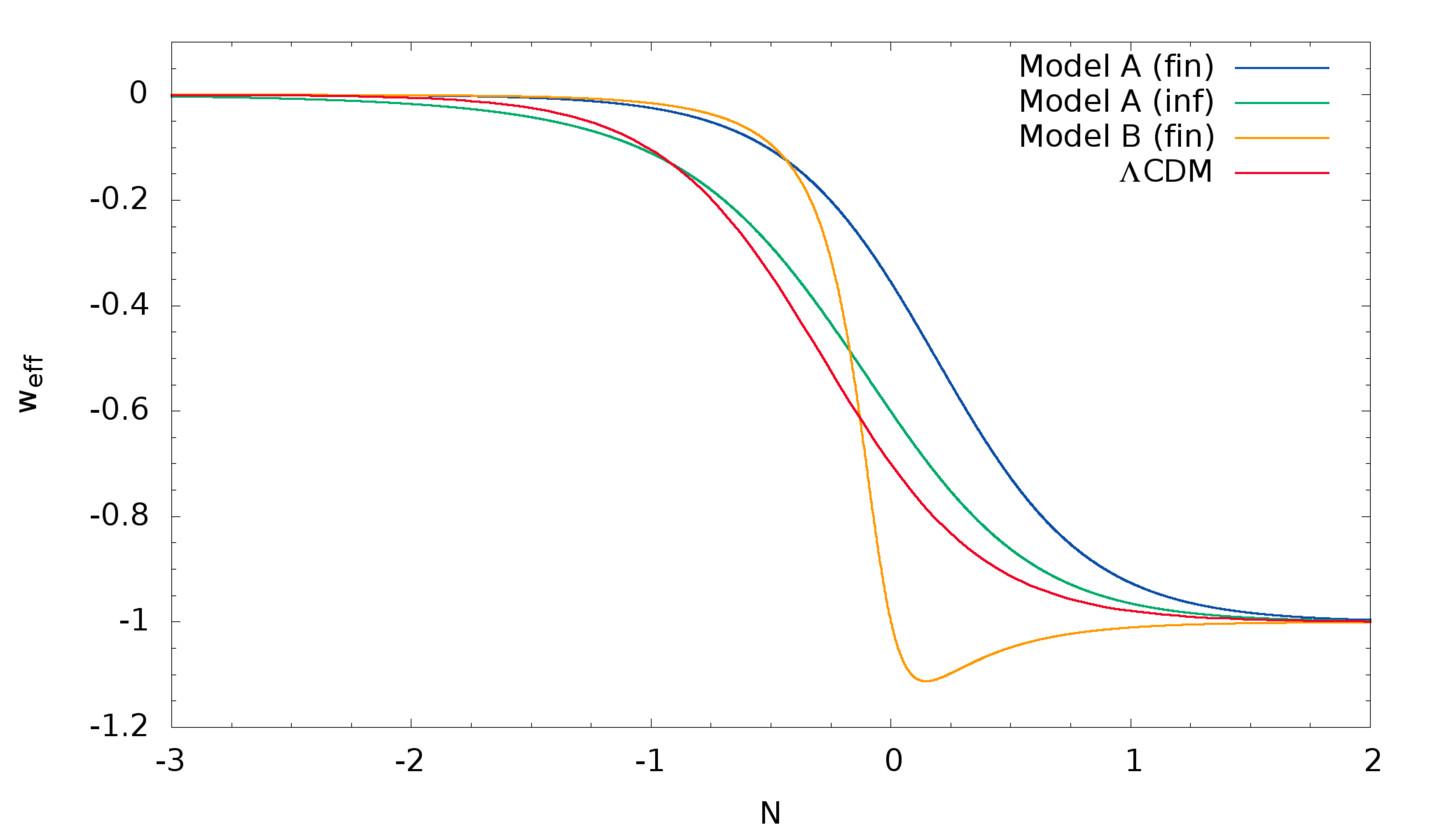}\includegraphics[width=0.45\textwidth]{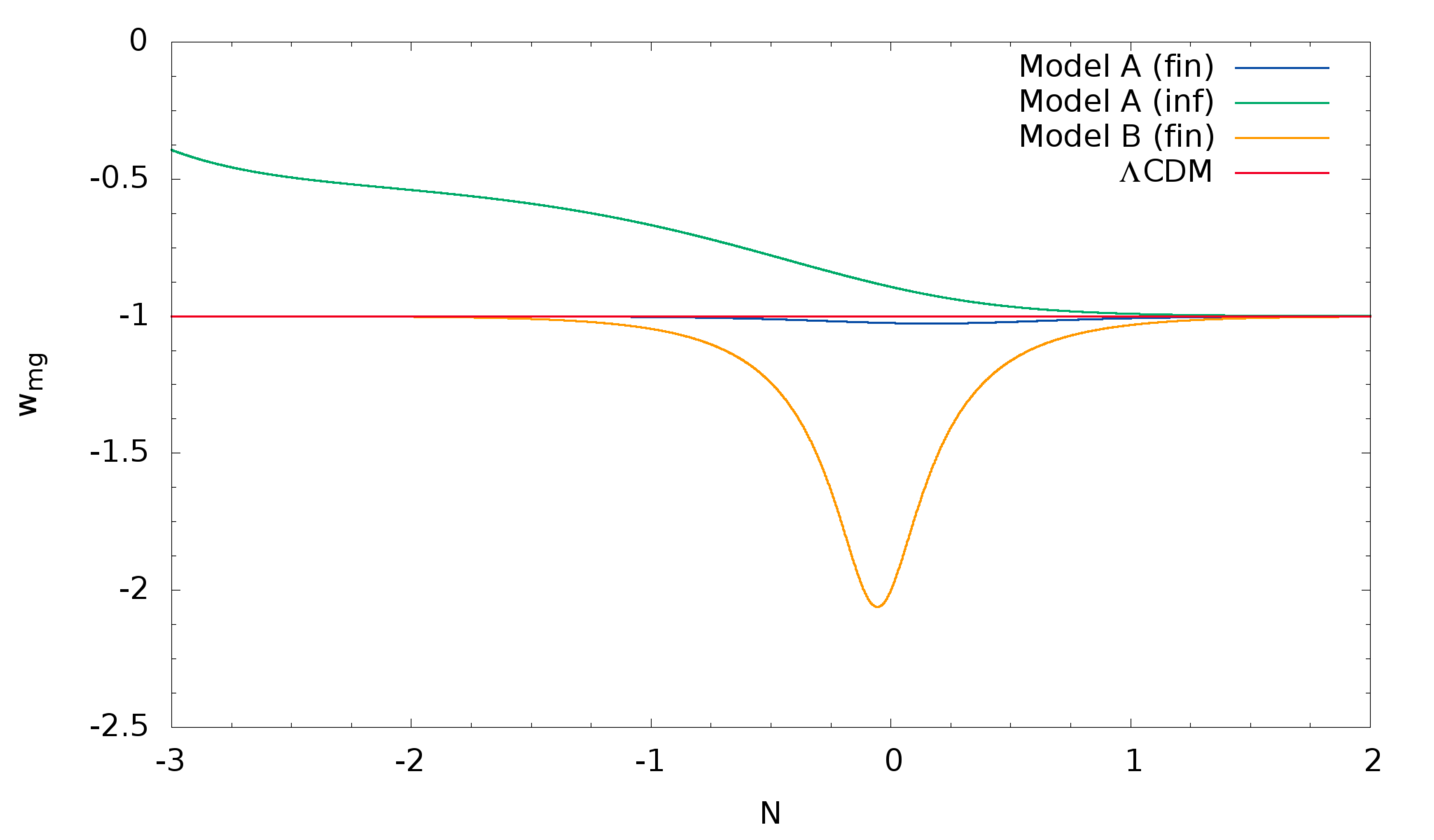}

\caption{A comparison of $r$ (top left), $\Omega_{m}$ (top right), $w_{eff}$
(bottom left) and $w_{mg}$ (bottom right) of all viable branches
in the models A (blue and green) and B (yellow) whose evolution of
$r'$ were already discussed in Figure \ref{fig:example_models_rp}.
Additionally, the latter three plots contain the $\Lambda$CDM expectation
for $\Omega_{m}=0.3$. \label{fig:example_models_cosmquantities}}
\end{figure}

\textcolor{black}{We chose two representative models to sketch a possible
viable evolution of a bimetric gravity in Figures \ref{fig:example_models_rp}
and \ref{fig:example_models_cosmquantities}. The model A, described
by $\beta_{i}=(1,\frac{1}{5},0,0,1)$, produces tw}o viable branches.
Although $\Omega_{m}$ and $w_{eff}$ evolve similarly in both branches,
we find a phantom equation of state only in the finite one. An one-parameter
model $\beta_{0}=\beta_{i}=\hat{\beta}$, such as model B with $\hat{\beta}=\frac{4}{9}$,
is only able to produce a viable finite branch. Those models always
produce a phantom since a positive expansion rate requires $\hat{\beta}>0$.

\section{\textcolor{black}{Comparing to Supernovae Ia Hubble diagram}}

\textcolor{black}{To compare the background evolution of bimetric
gravity models with observed SNe Ia, we use the SCP Union 2.1 Compilation
\cite{Suzuki:2011hu} containing 580 SNe Ia. For each observed SN
Ia we can use the measured maximum magnitude in the B-band \ensuremath{m_{B}^{\text{max}}}
together with the stretch correction \ensuremath{s} and the color
correction \ensuremath{c} to compute the likelihood for a bimetric
model \ensuremath{\theta} with 
\begin{equation}
L(\theta)\propto\int\exp\left(-\sum_{i=1}^{N}\frac{(\mu_{i}-\mu_{\text{theo}})^{2}}{2\sigma_{i}^{2}}\right)\text{d}M\,\text{d}\alpha\,\text{d}\beta\;,
\end{equation}
where \ensuremath{\alpha} and \ensuremath{\beta} are nuisance parameters
which weight the stretch- and color correction and \ensuremath{M}
denotes the absolute magnitude, 
\begin{equation}
\mu_{i}=m_{B_{i}}^{\text{max}}-M+\alpha\left(s_{i}+1\right)-\beta c_{i}\;.
\end{equation}
The marginalization over \ensuremath{M} can be performed analytically,
whereas we simplify the computation by using the values for \ensuremath{\alpha}
and \ensuremath{\beta} that minimize \ensuremath{\chi_{\text{red}}^{2}}
instead of computing the marginalization numerically. In addition,
we add an intrinsic dispersion which we assume to be \ensuremath{\sigma_{\text{int}}=0.1345\,\text{mag}}
in order to obtain a $\chi_{\text{red}}^{2}=1$ for the best fit in
a $\Lambda$CDM cosmology.}

We decided not to use other cosmological datasets like baryon acoustic
oscillations because they are at the moment far weaker than SN Ia
and CMB peak positions because their analysis depends on various assumptions
which are not warranted in a non-standard model as the one we explore
here. Nevertheless, our results are in agreement with Ref. \cite{2013JHEP...03..099A},
where these additional datasets have been employed.

\section{\textcolor{black}{Minimal models: 1-parameter models}}

\textcolor{black}{It is very instructive to study in detail some simple
subset among all the possible viable cosmologies. During our analysis
we will mostly assume a vanishing explicit cosmological constant,
i.e. $\beta_{0}=0$ (the only model with a non-vanishing cosmological
constant that is studied in this work will be the 1-parameter model
$\beta_{0}=\beta_{i}=\hat{\beta}$). This subset of models is a very
interesting one since those models may fit observational data without
the need of a  cosmological constant. In this section we assume moreover
that all of the other $\beta_{i}$ vanish, except one, i.e. we restrict
ourselves to $\beta_{i}$ models. In this case, as already shown,
only one possibility, the $\beta_{1}$ model, turns out to be viable.
In terms of simplicity, this is the minimal bimetric model, so it
can help us gaining intuition on the behavior of this class of models.
This model was already studied and compared to the SN Ia data in \cite{2013JHEP...03..099A};
the same paper excludes the other $\beta$ models on the ground of
their poor fit to data. The $\beta_{1}$ model is interesting also
because $r$ can be easily solve}d analytically. Its evolution follows
from Eq. (\ref{eq:rprime}), 
\begin{equation}
r'=\frac{3r\left(1-3r^{2}\right)}{1+3r^{2}}\;.
\end{equation}
\textcolor{black}{Note that the evolution of $r$ does not depend
on $\beta_{1}$. In terms of the scale factor, the solution r}eads
\begin{equation}
r(a)=\frac{1}{6}a^{-3}\left(-A\pm\sqrt{12a^{6}+A^{2}}\right)\;.\label{eq:minModel_r}
\end{equation}
To determine the constant $A$, we use the background equation Eq.
(\ref{eq:fried-1}) at current time which provides $r_{0}=\frac{1-\Omega_{m0}}{\beta_{1}}$
and therefore 
\begin{equation}
A=\frac{3(\Omega_{m0}-1)^{2}-\beta_{1}^{2}}{\beta_{1}\left(\Omega_{m0}-1\right)}\;.\label{eq:minModel_A}
\end{equation}
Depending on $\beta_{1}$ and $\Omega_{m0}$, both a negative and
positive $A$ is possible. To satisfy $r(a\rightarrow0)=0$, we need
to choose the positive sign in Eq. (\ref{eq:minModel_r}) if $A$
is positive, or the negative sign in case of a negative $A$. The
comparison with the SNIa Hubble diagram shows that $A$ has to be
positive (see below).

With this result, the equation of state and $\Omega_{m}$ are fully
described through 
\begin{align}
\Omega_{m}(a) & =-\frac{1}{6}Aa^{-6}\left(A\mp\sqrt{12a^{6}+A^{2}}\right)\;,\\
w_{eff}(a) & =\pm\frac{A}{\sqrt{12a^{6}+A^{2}}}-1\;,\\
w_{mg}(a) & =\mp\frac{A}{\sqrt{12a^{6}+A^{2}}}-1\;.
\end{align}
T\textcolor{black}{hus, in the $\beta_{1}$ viable minimal model,
the equation of state always evolves from $-2$ to $-1$. These equations
imply a simple and testable relation between $w_{mg}$ and $\Omega_{m}$
valid at all times during matter domination: 
\begin{equation}
w_{mg}=\frac{2}{\Omega_{m}-2}\;.\label{eq:testable}
\end{equation}
}

\textcolor{black}{In general, denoting with a subscript $0$ the present
time, the following conditions must be satisfied by any model:}

\begin{align}
\Omega_{m0} & =1-\frac{B_{0}(r_{0})}{B_{1}(r_{0})}r_{0}\;,\\
1 & =\frac{B_{1}(r_{0})}{3r_{0}}\label{eq:system}
\end{align}
(the last one is obtained from Eq. (\ref{eq:frif}) after expressing
the $\beta$s in units of $H_{0}^{2}$). In particular, for the $\beta_{1}$
minimal model we obtain then a direct relation to the present value
of the matter fractional density, $\beta_{1}=\sqrt{3(1-\Omega_{m0})}$
which yields 
\begin{equation}
A=\frac{\sqrt{3}\Omega_{m0}}{\sqrt{1-\Omega_{m0}}}\text{\;.}
\end{equation}

\begin{figure}
\includegraphics[width=0.45\textwidth]{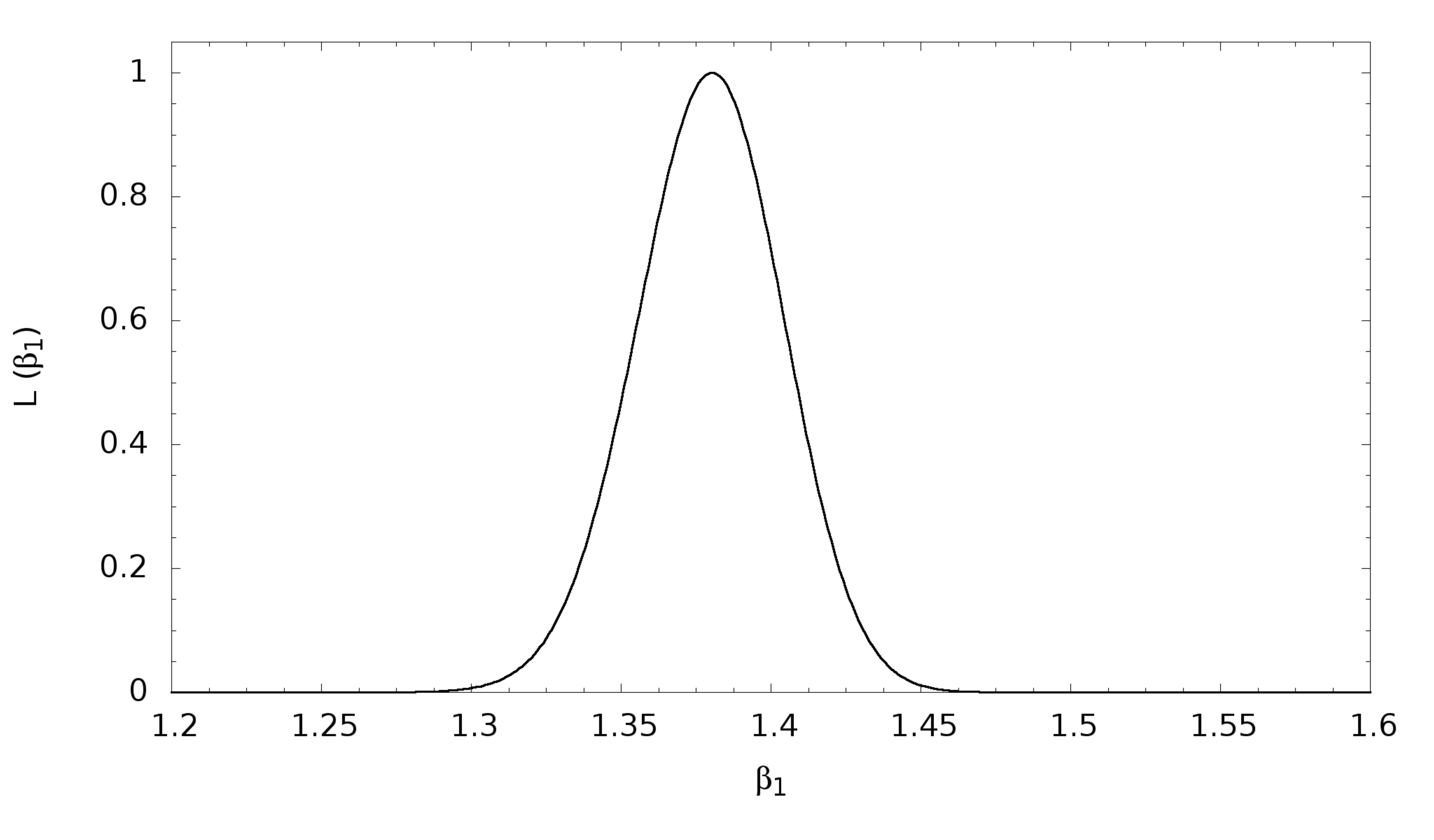}\includegraphics[width=0.45\textwidth]{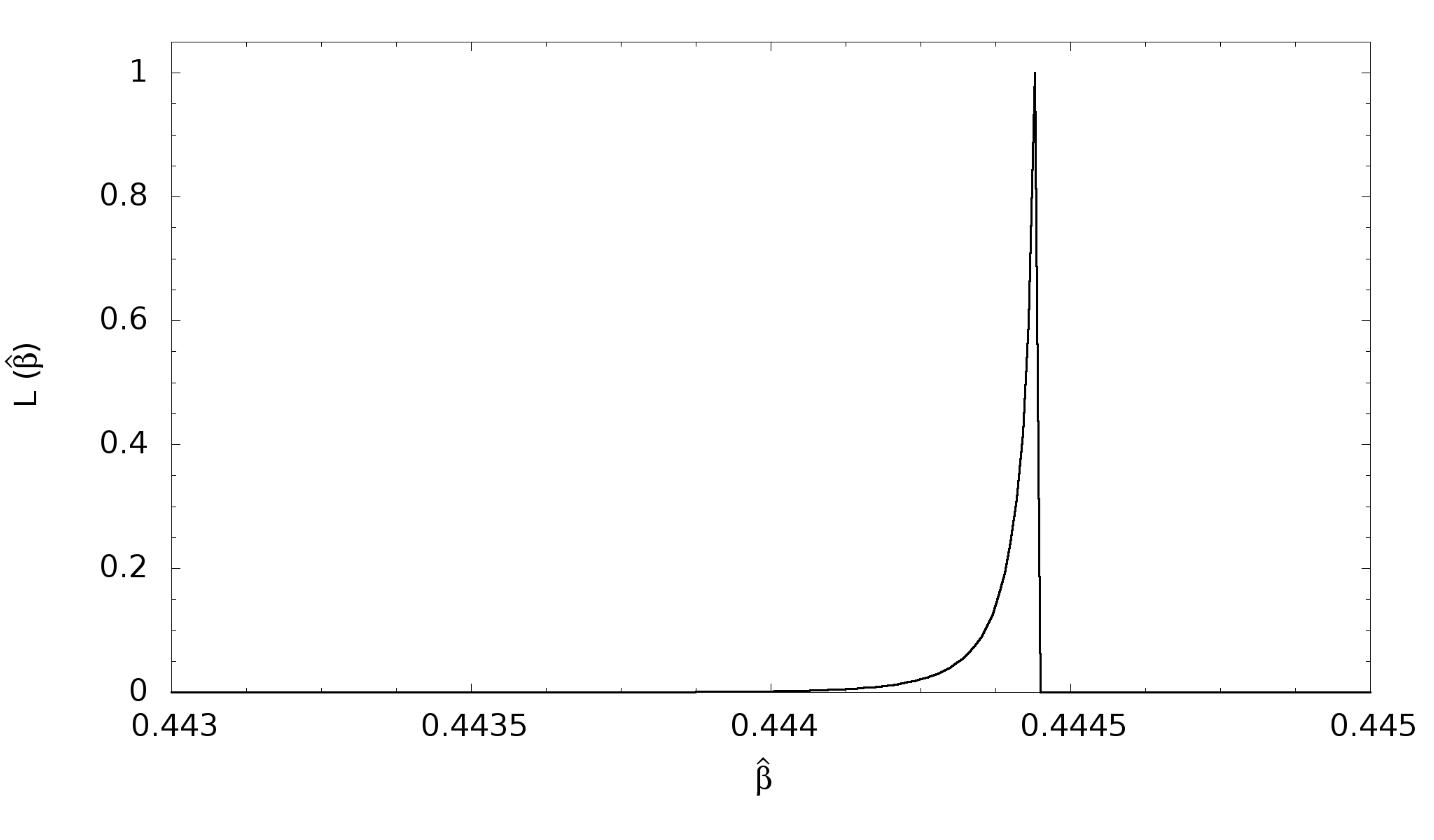}

\caption{Likelihood for the coupling parameter in the minimal $\beta_{1}$
(left) and $\hat{\beta}$ model (right). The maxima of the likelihoods
were rescaled to unity. Note that the $\hat{\beta}$ model produces
non-viable solutions for $\hat{\beta}>\frac{4}{9}$. \label{fig:likelihood_minmodel}}
\end{figure}

We fitted the $\beta_{1}$ model to the SN Union 2.1 catalog (see
Figure \ref{fig:likelihood_minmodel}) and obtained $A\approx0.8$
for the best fit. The most likely values for $\beta_{1}$ and $\Omega_{m0}$
are s\textcolor{black}{ummarized in Table \ref{tab:bei-models}.{}
We list also the pr}esent value of the equation of state expressed
using the simple CPL parametrization \cite{2001IJMPD..10..213C,2003PhRvL..90i1301L}
\begin{equation}
w(a)=w_{0}+w_{a}(1-a)
\end{equation}
\textcolor{black}{in order to provide a quick comparison to present
and future cosmological data.}

\textcolor{black}{The $\beta_{1}$ model is then a valid alternative
to $\Lambda$CDM in terms of simplicity, and although it does not
reduce to $\Lambda$CDM in any limit, it gives a good fit to the background
data.}

A second type of minimal models is described by identical couplings
$\beta_{0}=\beta_{i}=\hat{\beta}$. As noted earlier, only those models
with $0<\hat{\beta}\leq\frac{4}{9}$ produce one viable finite branch.
The evolution of $r$, described by 
\begin{equation}
r'=\frac{3r\left(1-r^{2}\right)}{1-2r+3r^{2}}\;,
\end{equation}
has an analytical solution, though it is much more complicated than
in a minimal model with only one non-vanishing coupling. However,
the matter density parameter follows the simple relation 
\begin{equation}
\Omega_{m}=1-r
\end{equation}
which, just like $r'$, is independent of $\hat{\beta}$. Of course,
the present value $r_{0}$ is a function of $\hat{\beta}$. Again,
we can use the set of equations (\ref{eq:system}) to obtain a relation
between $\hat{\beta}$ and $\Omega_{m0}$, 
\begin{equation}
\hat{\beta}=\frac{3\left(\Omega_{m0}-1\right)}{\left(\Omega_{m0}-2\right)^{3}}\;.
\end{equation}

We found that both types of minimal models are only able to produce
viable branches if the coupling parameters are positive and $r_{0}$
is located in a finite branch. Then Eq. (\ref{eq:wmg}) directly implies
that all these minimal models are described by a phantom equation
of state at any time. A comparison of both minimal models with observed
SNe Ia yields the likelihoods in Figure \ref{fig:likelihood_minmodel}
which provide the best fits listed in Table \ref{tab:minimal_models_bestfit}.
Their equation of state is plotted in Figure \ref{fig:bestfit_wmg_weff}.

\begin{table}
\begin{centering}
\begin{tabular}{|c|c|c|c|c|c|c|}
\hline 
 & $\beta(\Omega_{m0})$  & $\chi_{\text{{min}}}^{2}$  & $\beta_{1}$ or $\hat{\beta}$  & $\Omega_{m0}$  & $w_{0}$  & $w_{a}$\tabularnewline
\hline 
\hline 
$\beta_{1}$  & $\sqrt{3(1-\Omega_{m0})}$  & $578.3$  & $1.38_{-0.03}^{+0.03}$  & $0.37_{-0.02}^{+0.02}$  & $-1.22_{-0.02}^{+0.02}$  & $-0.64_{-0.04}^{+0.05}$\tabularnewline
\hline 
$\hat{\beta}$  & $\frac{3(\Omega_{m0}-1)}{(\Omega_{m0}-2)^{3}}$  & $606.3$  & $0.44_{-0.01}^{+0.00}$  & $0.50{}_{-0.00}^{+0.01}$  & $-2.00{}_{-0.01}^{+0.00}$  & $-1.97{}_{-0.00}^{+0.07}$\tabularnewline
\hline 
\end{tabular}
\par\end{centering}

\caption{Best fit values for the two minimal models. The column $\beta(\Omega_{m0})$
lists the relation between the value of the coupling parameter $\beta_{1}$
and $\hat{\beta}$, respectively, and the present matter density parameter.
The parameters $w_{0}$ and $w_{a}$ describe the CPL parametrization
at present time. \label{tab:bei-models} \label{tab:minimal_models_bestfit}}
\end{table}

\section{Two-parameter models}

We move now to models in which all $\beta_{i}$ vanish except two,
taken in turn to be all possible combinations (we keep $\beta_{0}=0$).
As already shown, we need to exclude all cases in which $\beta_{1}=0$
since we do not expect any viable models.

\begin{figure}
\includegraphics[height=5cm]{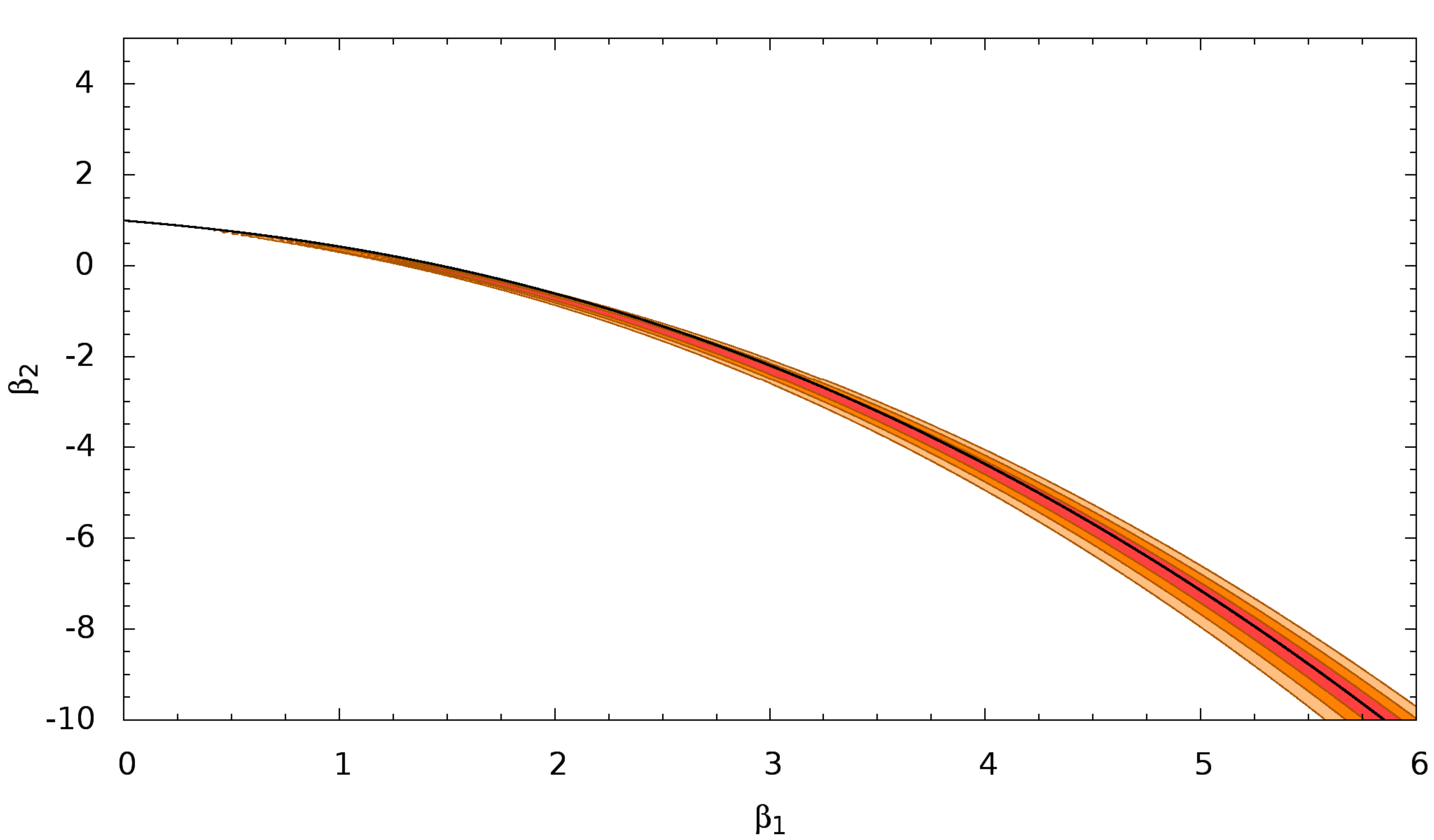}\includegraphics[height=5cm]{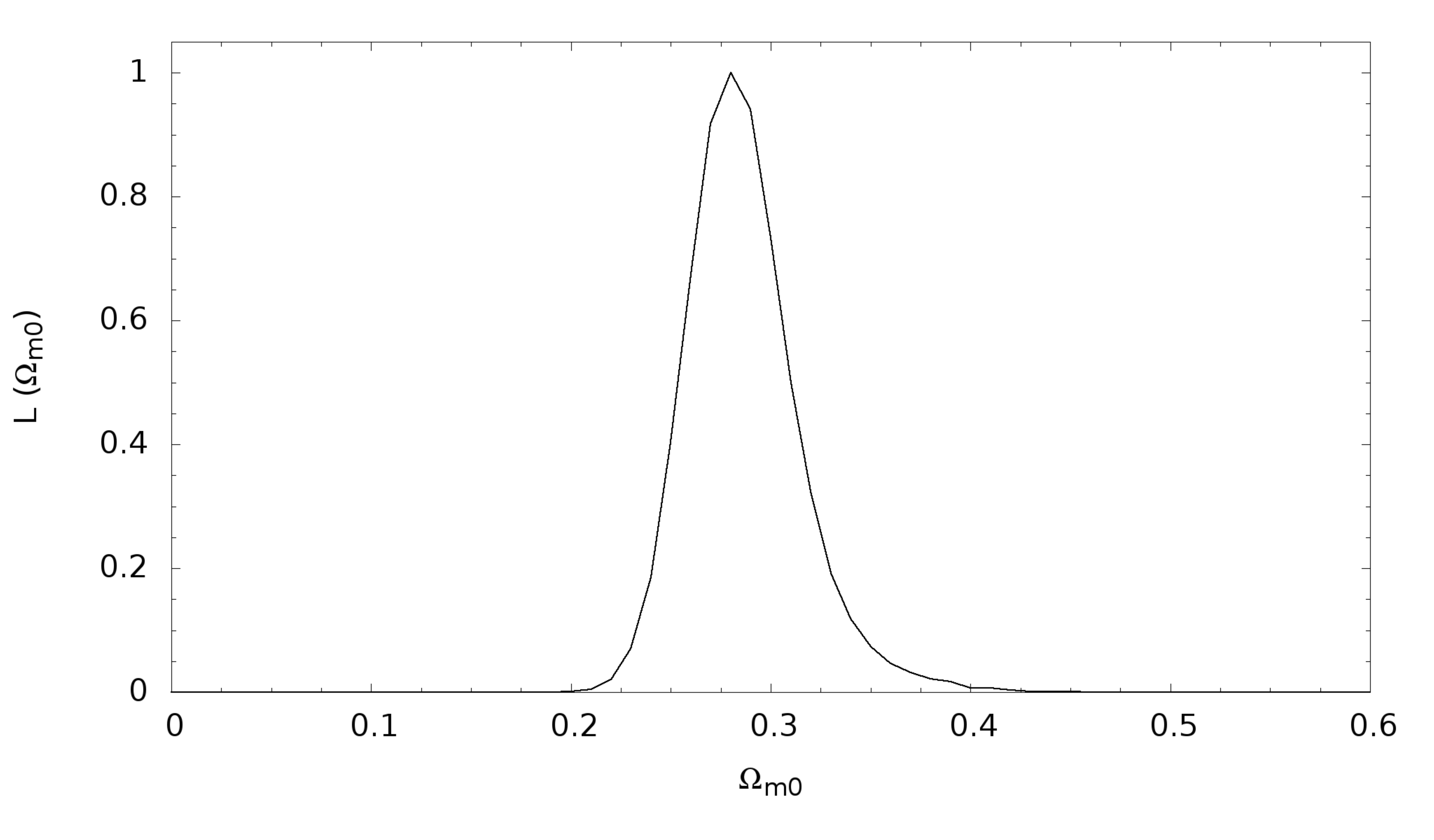}

\includegraphics[height=5cm]{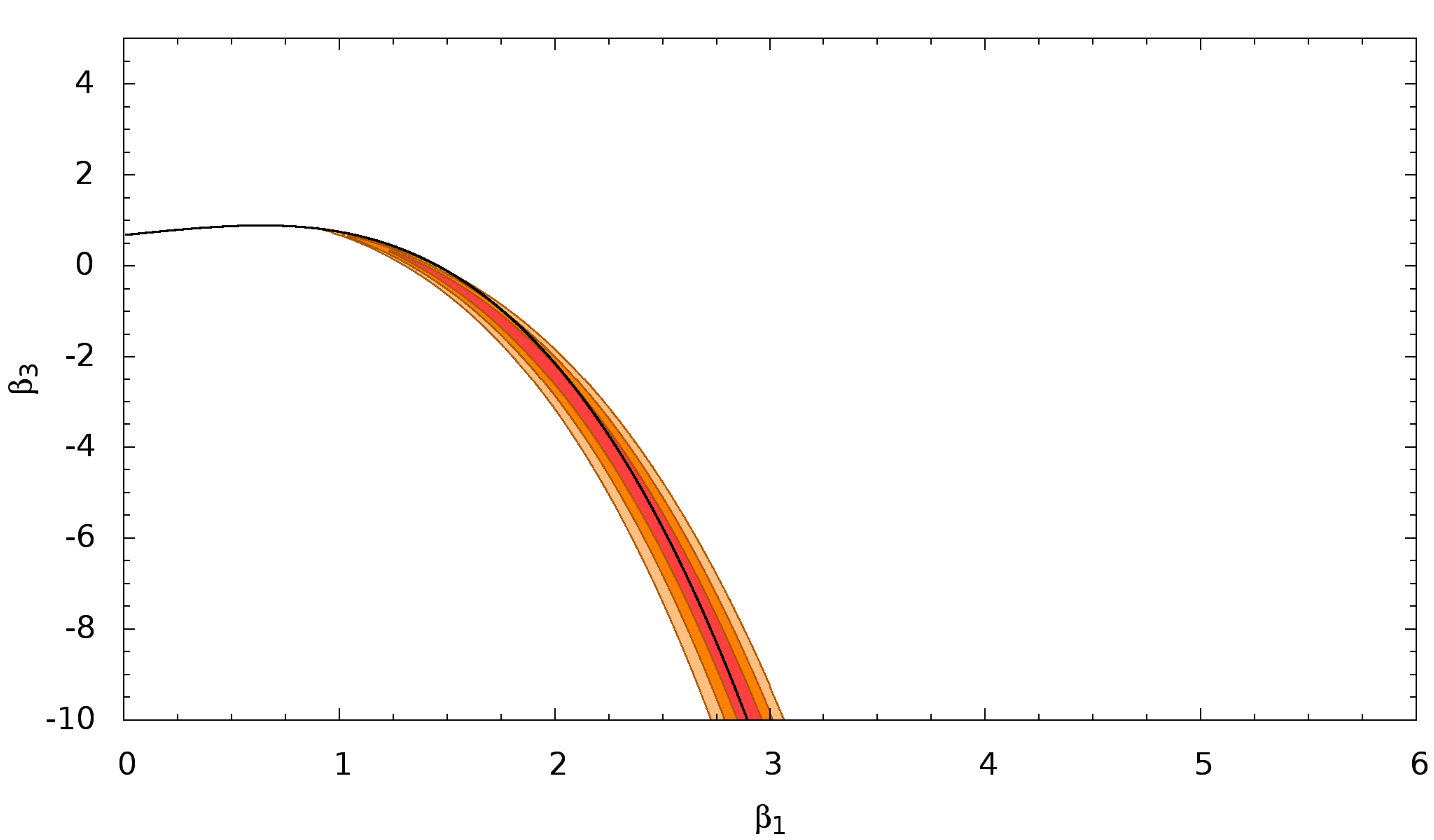}\includegraphics[height=5cm]{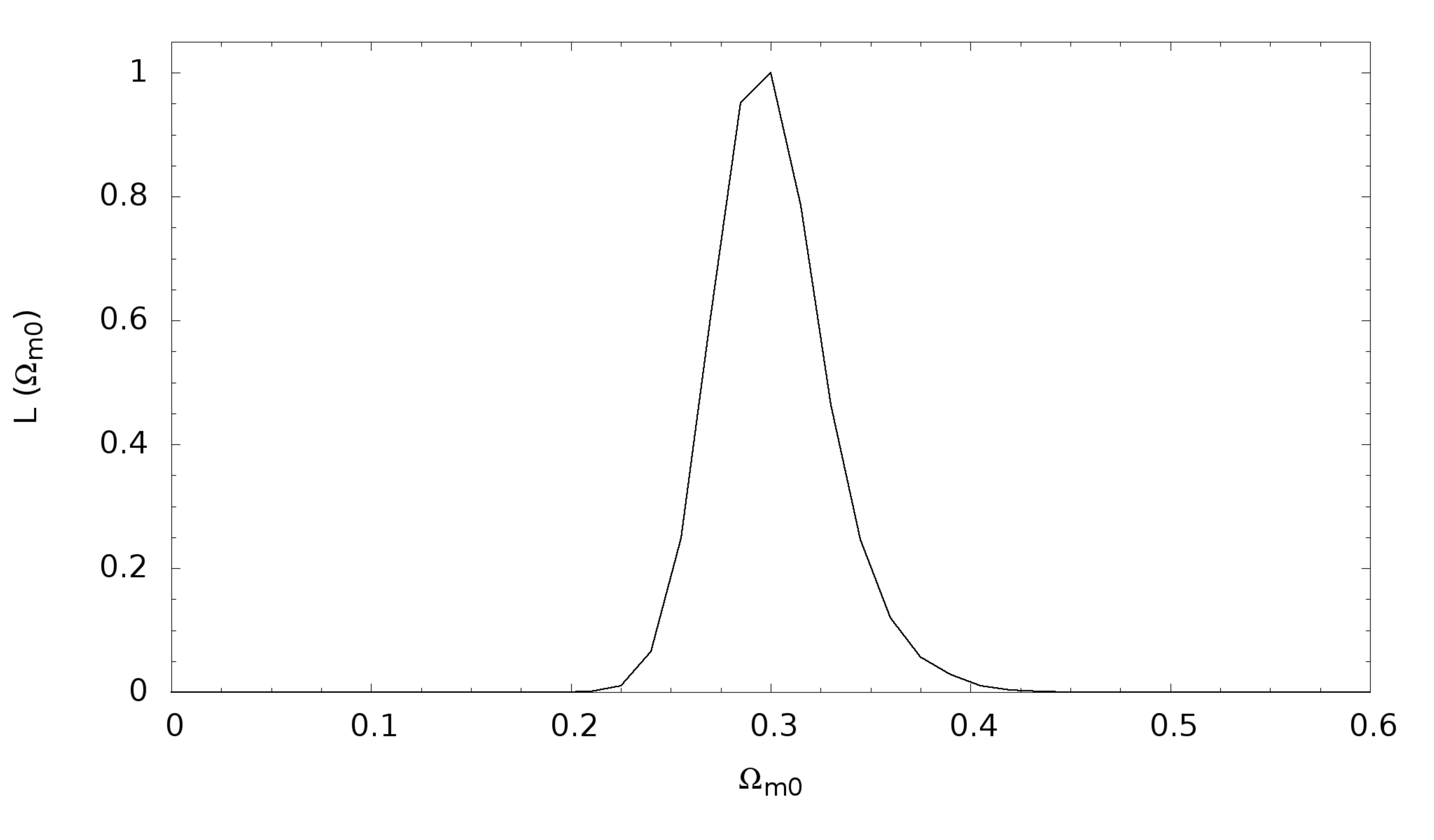}

\includegraphics[height=5cm]{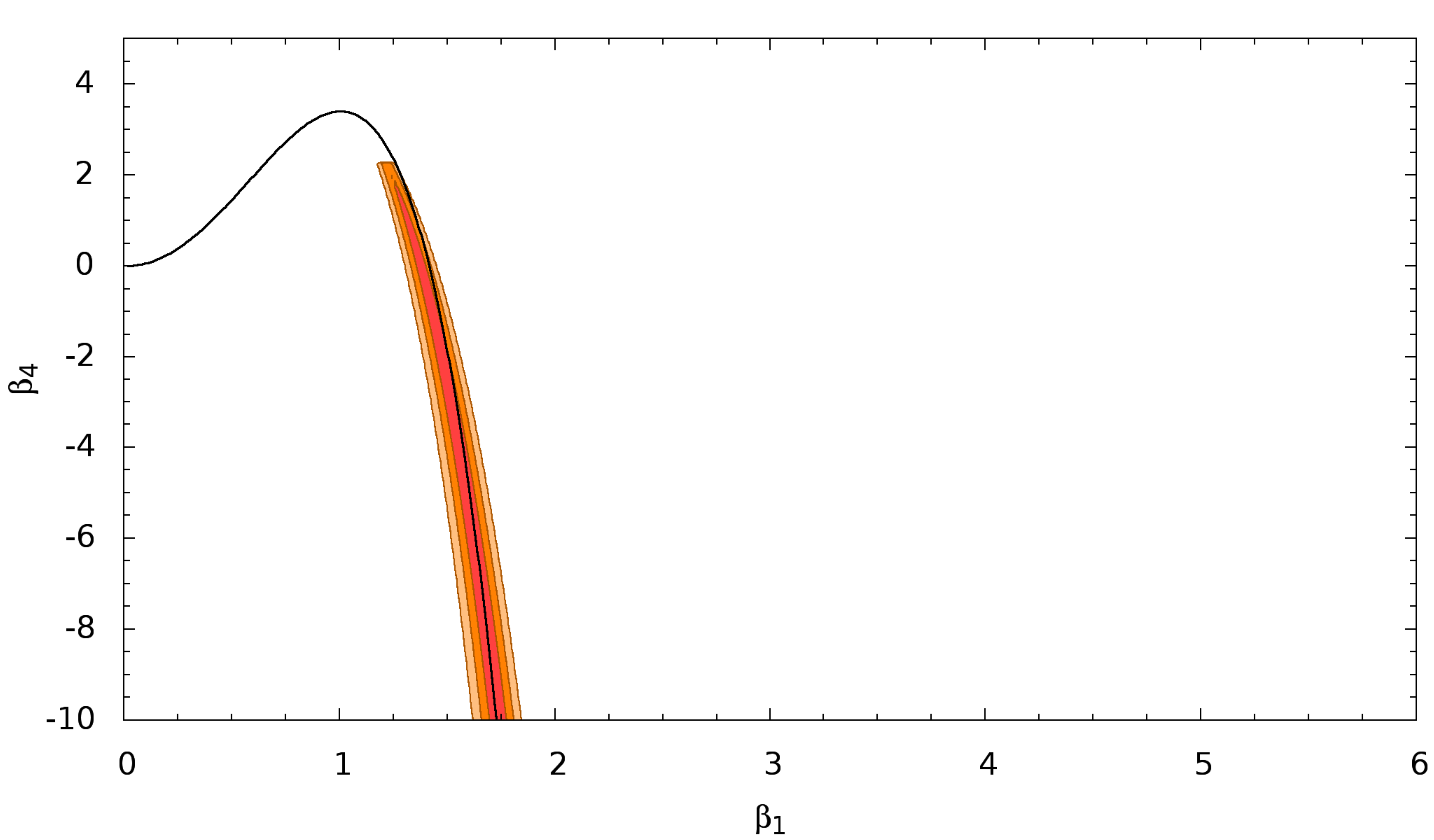}\includegraphics[height=5cm]{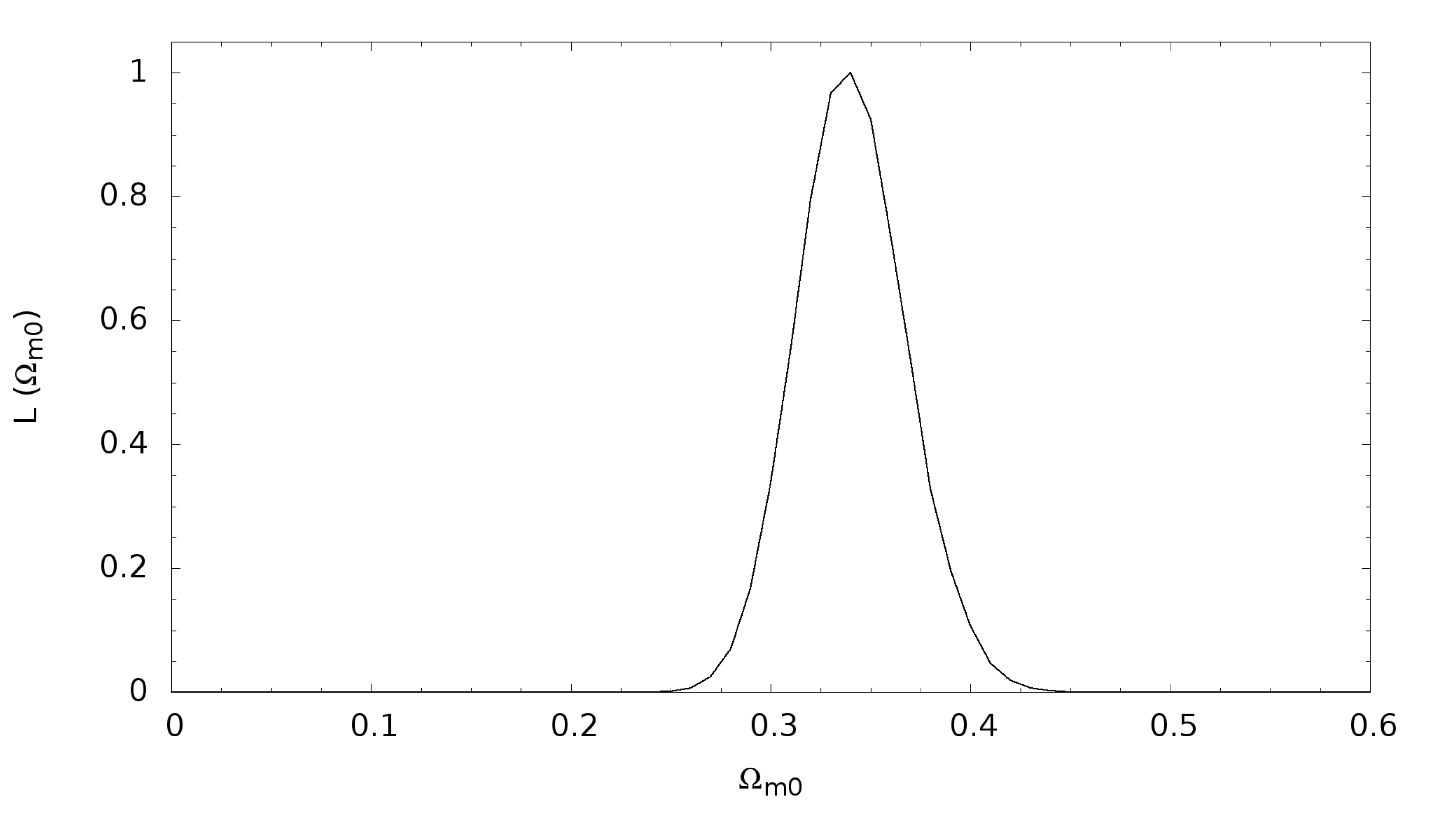}

\includegraphics[height=5cm]{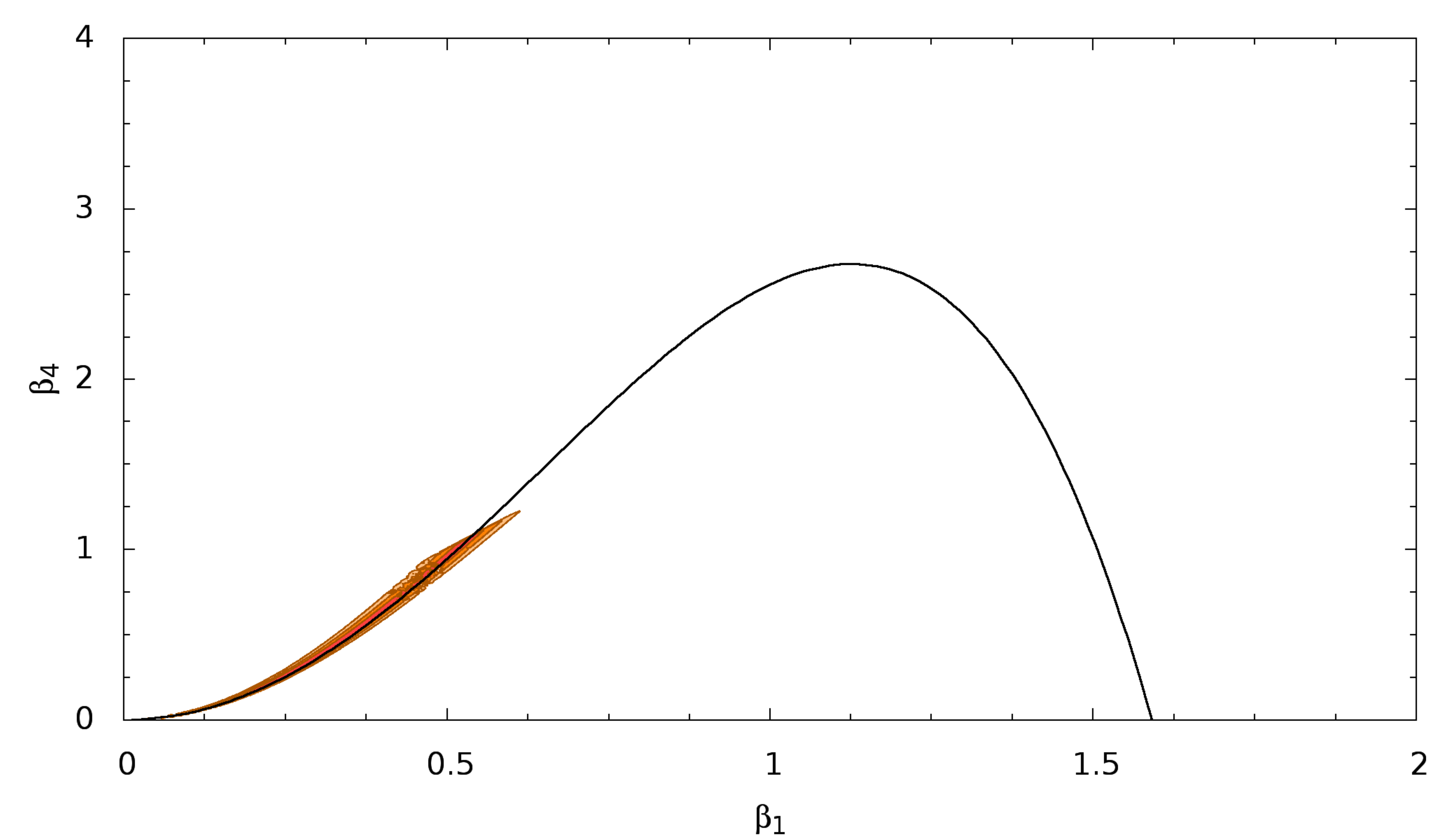}\includegraphics[height=5cm]{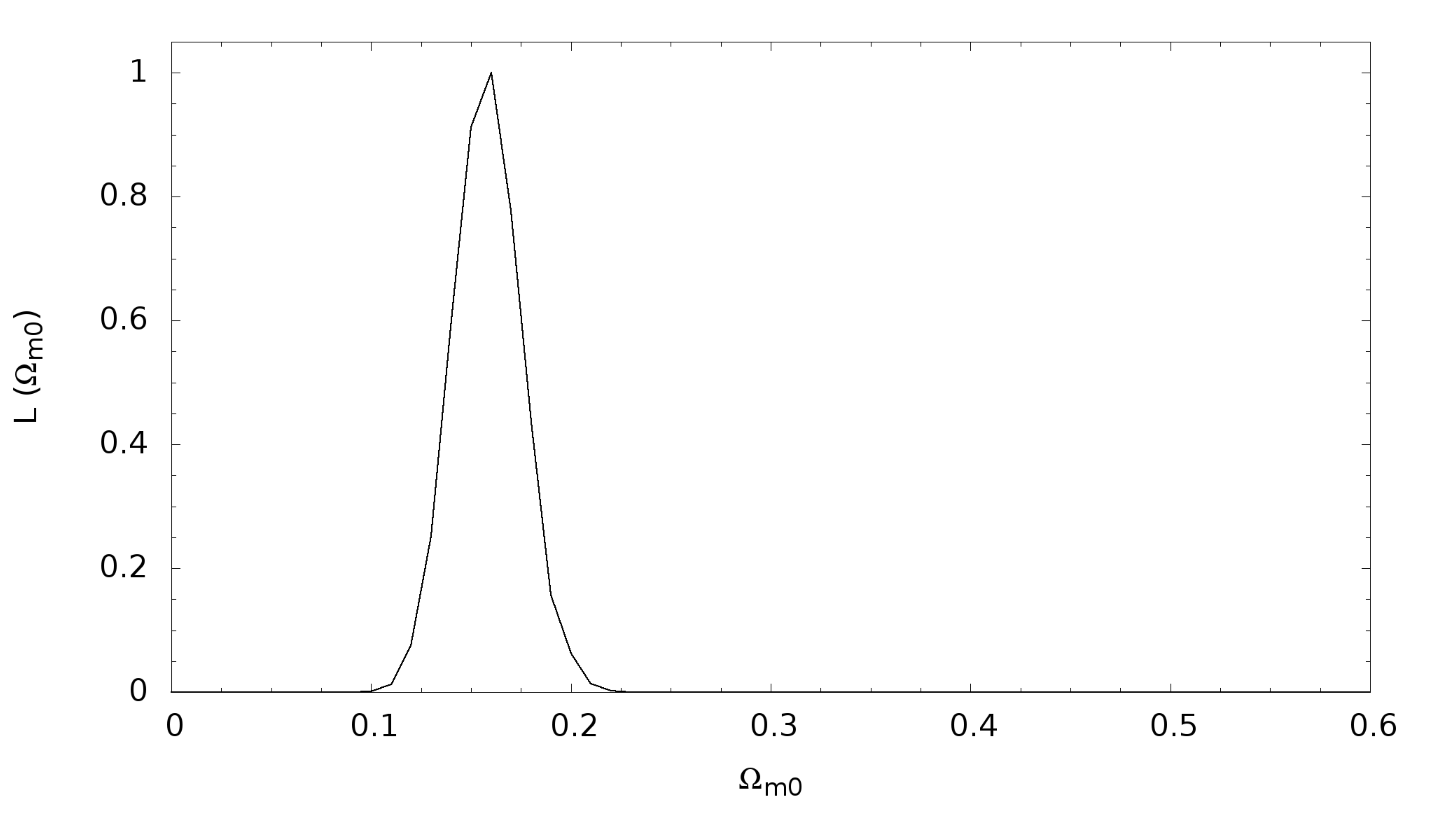}

\caption{Left: Likelihoods from observed SNe Ia with only two $\ensuremath{\beta}$-parameter
varying while all other $\ensuremath{\beta_{i}}$ vanish.\textcolor{red}{{}
}In $\beta_{1}\beta_{4}$ models we distinguish between finite (plots
in third row) and infinite (last row plots) branches. The filled regions
correspond to the 68\% (red), 95\% (orange) and 99.7\% (yellow) confidence
level. In each two-dimensional likelihood, the analytic result $\beta_{j}(\beta_{i},\Omega_{m0})$
is illustrated by a black solid line and corresponds to the most likely
value $\Omega_{m0}$. Right: Likelihood for $\Omega_{m,0}$ obtained
after a marginalization over the $\beta$ parameters corresponding
to the likelihoods on the left side.\label{fig:SNeIa_likelihoods} }
\end{figure}

To compute the likelihood for \ensuremath{\Omega_{m,0}}, we divide
the range in \ensuremath{\Omega_{m,0}} in bins \ensuremath{B_{k}}
of constant width and marginalize the likelihood over both \ensuremath{\beta}-parameters
with the restriction \ensuremath{\Omega_{m,0}\in B_{k}}. Our results
are summarized in Figure \ref{fig:SNeIa_likelihoods} where the left
plots show the 68\%, 95\% and 99.7\% confidence regions in the \ensuremath{\beta_{i}-\beta_{j}}
plane, the corresponding likelihoods for \ensuremath{\Omega_{m,0}}
are illustrated in the right column. In all cases that are shown in
Figure \ref{fig:SNeIa_likelihoods}, we found bimetric gravity models
which are consistent with observed SNe Ia. We always observe a strong
degeneracy between the two free parameters, as already remarked in
Ref. \cite{2013JHEP...03..099A}.

As in the minimal cases, the system (\ref{eq:system}) gives a relation
between pairs of $\beta$ and $\Omega_{m0}$: 
\begin{enumerate}
\item For $\beta_{1}\beta_{2}$: 
\begin{align}
\beta_{2} & =\frac{\beta_{1}^{2}+\sqrt{\beta_{1}^{4}-9\beta_{1}^{2}\Omega_{m0}+9\beta_{1}^{2}}}{9\left(\Omega_{m0}-1\right)}+1\;.
\end{align}

\item For $\beta_{1}\beta_{3}$: 
\begin{align}
\beta_{3} & =\frac{-32\beta_{1}^{3}\pm\sqrt{\left(8\beta_{1}^{2}+27\left(\Omega_{m0}-1\right)\right)^{2}\left(16\beta_{1}^{2}-27\left(\Omega_{m0}-1\right)\right)}-81\text{\ensuremath{\beta_{1}}}\left(\Omega_{m0}-1\right)}{243\left(\Omega_{m0}-1\right)^{2}}\;,
\end{align}
where the positive sign should be taken if $\beta_{1}<\frac{3}{2}\sqrt{\frac{3}{2}}\sqrt{1-\Omega_{m0}}$
and the negative one otherwise. 
\end{enumerate}
In all cases $\Omega_{m0}$ should be taken as the best fit value.
The $\beta_{1}\beta_{4}$ model does not have a simple analytic solution
but the relation is easily solved numerically. These relation are
plotted in the same Fig. \ref{fig:SNeIa_likelihoods}; as one can
see, they fit very well the degeneracy curves.

\begin{figure}
\includegraphics[width=0.45\textwidth]{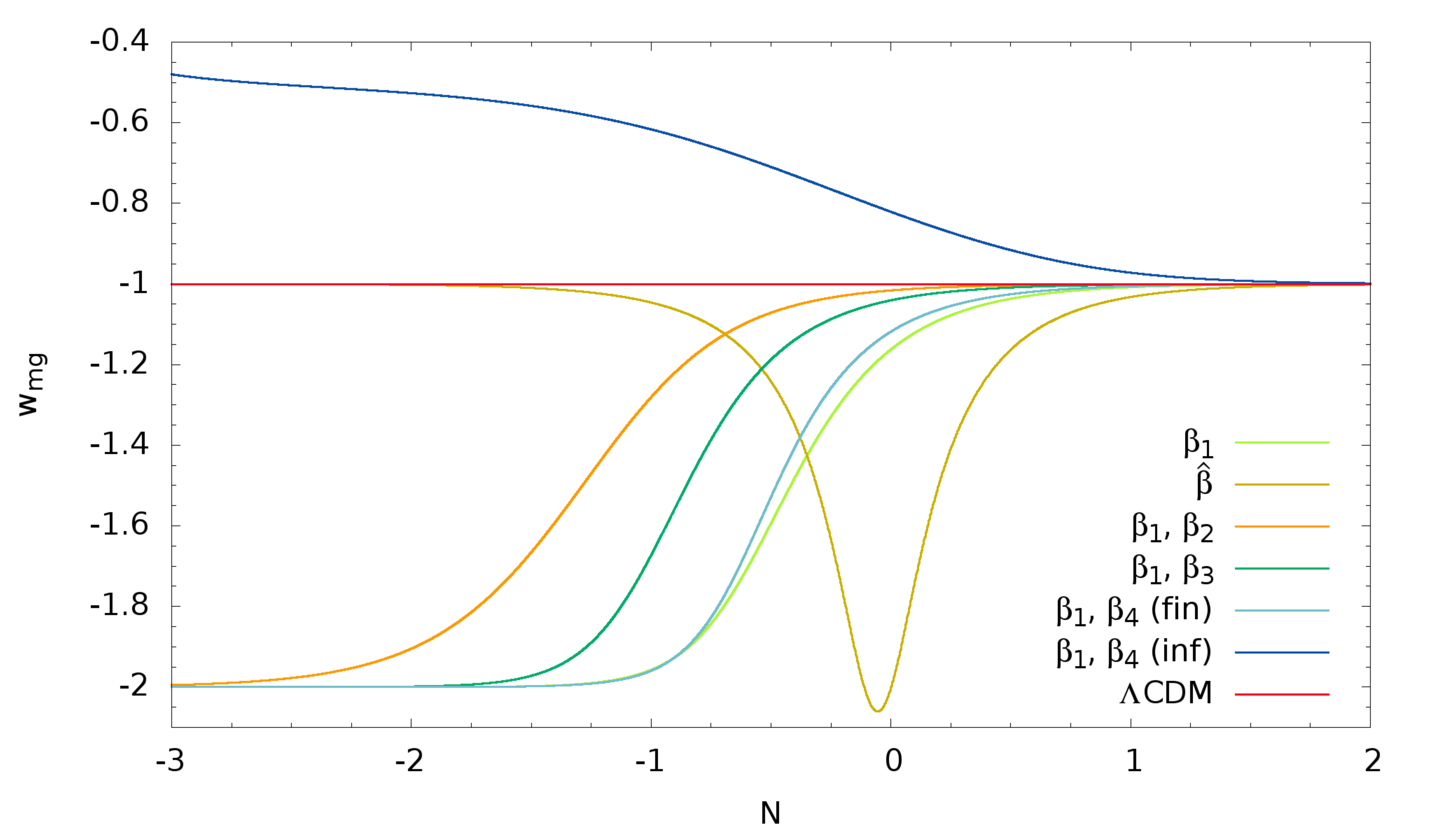}\includegraphics[width=0.45\textwidth]{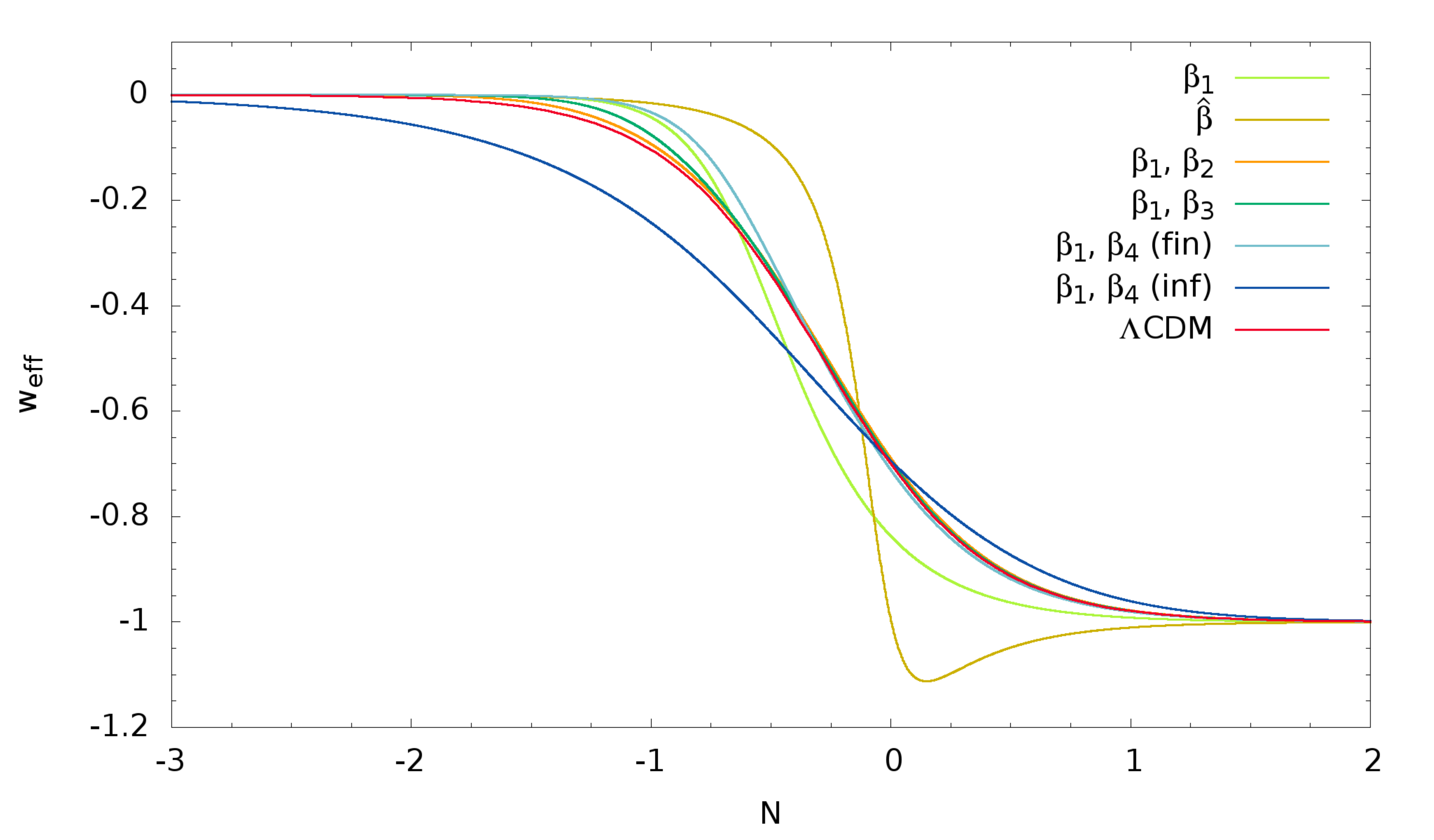}

\caption{Evolution of the equation of state in the best fits in the minimal
$\beta_{1}$ and $\hat{\beta}$ models and the two-parameter models
$\beta_{1}\beta_{2},$ $\beta_{1}\beta_{3}$ and $\beta_{1}\beta_{4}$.
Here, we distinguish between finite (light blue) and infinite (dark
blue) branches in $\beta_{1}\beta_{4}$ models. \label{fig:bestfit_wmg_weff}}
\end{figure}

\begin{figure}
\includegraphics[width=0.8\textwidth]{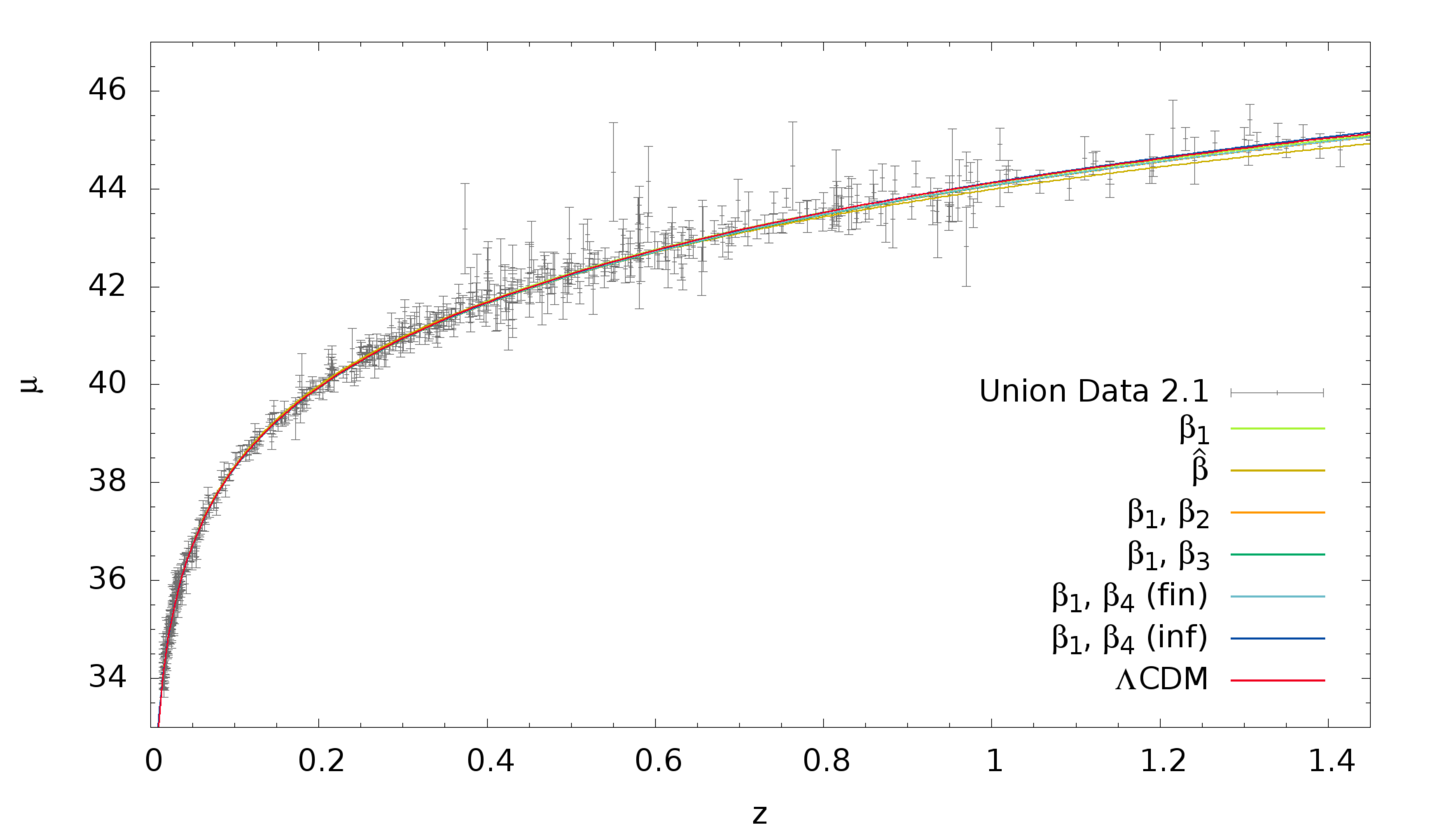}\caption{Hubble diagram with the best fit in the minimal one-parameter models
and two-parameter models compared to all measured SNe Ia from the
Union Data 2.1. As already indicated by the numerical values of the
$\chi^{2}$ (see Tables \ref{tab:bei-models} and \ref{table:fit_results}),
the best fit in the $\beta_{1}$ and in all analyzed two parameter
models are close to the $\Lambda$CDM result (red). \label{fig:distmod}}
\end{figure}

At $1\sigma$, the relative error $\Delta$ on the fitted $\beta_{j}(\beta_{i},\Omega_{m0})$,
with ($\beta_{i}<\beta_{j}$), are given in Table \ref{fig:SNeIa_likelihoods},
where we determined the error by fitting the 68\% contour with $\beta'_{j}=\beta_{j}(1+\Delta)$.
For the best fit in all analyzed combinations, we show the evolution
of the equation of state $w_{mg}$ in Figure \ref{fig:bestfit_wmg_weff}
and the distance moduli $\mu(z)$ in comparison with the measured
SNe Ia of the Union 2.1 Compilation in Figure \ref{fig:distmod}.

Note that the analytic fit does not always need to correspond to a
viable solution since it ignores the condition $0<r_{0}<r_{c}$ and
$r_{c}<r_{0}$ in the finite and infinite branch, respectively. We
therefore need to exclude some parameter regions. As an example, we
analyze all $\beta_{1}\beta_{3}$ models with positive $\beta_{3}$
and obtain 
\begin{equation}
r_{0}=\frac{3\pm\sqrt{9-12\beta_{1}\beta_{3}}}{6\beta_{3}}\quad\text{{and}\quad}r_{c}=\pm\sqrt{\frac{-3\left(\beta_{1}-\beta_{3}\right)\pm\sqrt{9\left(\beta_{1}^{2}+\beta_{3}^{2}\right)-14\beta_{1}\beta_{3}}}{2\beta_{3}}}\;.
\end{equation}
A necessary condition to satisfy the relation $0<r_{0}<r_{c}$ is
\begin{equation}
\beta_{3}<\frac{1}{243}\left(81\beta_{1}-32\beta_{1}^{3}+\sqrt{\left(27+16\beta_{1}^{2}\right)\left(-27+8\beta_{1}^{2}\right)^{2}}\right)
\end{equation}
which excludes most of the models with $\beta_{3}>0$. Similar boundaries
of the coupling parameter corresponding to the highest order interaction
exist in the $\beta_{1}\beta_{2}$ and $\beta_{1}\beta_{3}$ models,
too.

Only the model $\beta_{1}\beta_{4}$ is able to produces infinite
branches. The likelihoods in Figure \ref{fig:SNeIa_likelihoods} for
finite and infinite branches show that there is no parameter region
in which the contours of both likelihoods overlap. If there is a $\beta_{1}\beta_{4}$
model in which two viable branches co-exist, then at least one branch
is strongly disfavored by SNe Ia observations.

\begin{table}
\begin{tabular}{|c|c|c|c|}
\hline 
Model  & $\chi_{\text{{min}}}^{2}$  & $\Omega_{m0}$  & $\Delta$\tabularnewline
\hline 
\hline 
$\Lambda$CDM  & 578.00  & $0.27_{-0.02}^{+0.02}$  & \tabularnewline
\hline 
$\beta_{1},\beta_{2}$  & 577.99  & $0.28{}_{-0.03}^{+0.04}$  & $\sim0.03$\tabularnewline
\hline 
$\beta_{1},\beta_{3}$  & 578.02  & $0.30{}_{-0.04}^{+0.02}$  & $\sim0.08$\tabularnewline
\hline 
$\beta_{1},\beta_{4}$  & 578.04  & $0.34{}_{-0.04}^{+0.03}$  & $\sim0.20$\tabularnewline
\hline 
$\beta_{1},\beta_{4}$ (inf. branch $r\in(r_{c},\infty)$)  & 578.60  & $0.16{}_{-0.03}^{+0.02}$  & $\sim0.03$\tabularnewline
\hline 
\end{tabular}\caption{Numerical results of the best fit to SNe Ia data for different models
with only two free $\beta$-parameter. The relative error on the fit
$\beta_{j}(\beta_{i},\Omega_{m0})$ ($i<j$) corresponding to the
most likely value for $\Omega_{m0}$ is denoted by $\Delta$. \label{table:fit_results}}

\centering{} 
\end{table}

\section{conclusions}

\textcolor{black}{In this paper we studied a class of bimetric gravitational
models that have been shown to be ghost-free and to induce cosmological
acceleration. We define a viable cosmology as one in which the cosmic
evolution broadly resembles the standard one, without bounces, singularities
at finite time, and with a matter (or radiation) dominated past. Adopting
spatially flat metrics we find that the system becomes effectively
unidimensional and in some cases even analytical. This allows us to
find a number of simple rules for viability which selects a subset
of models and initial conditions. We show that if a branch is viable,
then its final state is always deSitter. We also find the analytical
condition for the occurrence of a phantom phase and we remark that
observing a phantom crossing would rule out the entire class of viable
bimetric models.}

\textcolor{black}{Then we show that among the models with only a single
non-zero parameter, only one gives a viable cosmology, which well
reproduces the SN data and can be taken as a simple, distinguishable
alternative to $\Lambda$CDM. The relation (\ref{eq:testable}) provides
a stringent test for this minimal model. For models with two coupling
constants} and without a cosmological constant,\textcolor{black}{{}
only three cases produce a viable cosmology. In several cases we find
also an analytic expression for the background best fit which very
closely approximates our numerical likelihood results. }

\textcolor{black}{These results allow to pre-select a number of cases
for which a detailed study, including perturbation growth, can be
performed. This task is carried out in a companion paper.}

\section*{Acknowledgment}

L.A. acknowledges support from DFG through the TRR33 project ``The
Dark Universe''. We are grateful to Y. Akrami, S. F. Hassan, S. Hoffmann,
T. Koivisto, D. Mota, M. Sandstand, A. Schmidt-May, M. von Strauss
for useful discussions. Special thanks to Valerio Marra for help with
the supernovae data and to Mariele Motta for checking various equations.
A.P. thanks the DAAD-WISE Fellowship for support.

\bibliographystyle{plainnat}  \bibliographystyle{plainnat}
\bibliography{observables,massive-gravity}

\end{document}